\documentstyle [12pt,eqsecnum,aps,amsfonts] {revtex}
\tighten
\draft
\input epsf
\newcommand{\beq}{\begin{equation}}
\newcommand{\eeq}{\end{equation}}
\newcommand{\beqs}{\begin{eqnarray}}
\newcommand{\eeqs}{\end{eqnarray}}

\newcommand{\gsim}{\mathrel{\raisebox{-.6ex}{$\stackrel{\textstyle>}{\sim}$}}}
\newcommand{\gtwid}{\mathrel{\raise.3ex\hbox{$>$\kern-.75em\lower1ex
\hbox{$\sim$}}}}
\newcommand{\ltwid}{\mathrel{\raise.3ex\hbox{$<$\kern-.75em\lower1ex
\hbox{$\sim$}}}}

\begin{document}

\begin{titlepage}

\begin{center}

{\Large \bf Complex-Temperature Phase Diagrams for the $q$-State Potts Model on
Self-Dual Families of Graphs and the Nature of the $q \to \infty$ Limit}

\vspace{1.1cm}
{\large Shu-Chiuan Chang${}$\footnote{email:
shu-chiuan.chang@sunysb.edu}
and Robert Shrock${}$\footnote{email:
robert.shrock@sunysb.edu }}\\
\vspace{18pt}
 C. N. Yang Institute for Theoretical Physics \\
 State University of New York \\
 Stony Brook, NY 11794-3840 \\
\end{center}

\vskip 0.6 cm

\begin{abstract}
\vspace{2cm}

We present exact calculations of the Potts model partition function $Z(G,q,v)$
for arbitrary $q$ and temperature-like variable $v$ on self-dual strip graphs
$G$ of the square lattice with fixed width $L_y$ and arbitrarily great length
$L_x$ with two types of boundary conditions.  Letting $L_x \to \infty$, we
compute the resultant free energy and complex-temperature phase diagram,
including the locus ${\cal B}$ where the free energy is nonanalytic.  Results
are analyzed for widths $L_y=1,2,3$.  We use these results to study the
approach to the large-$q$ limit of ${\cal B}$.



\end{abstract}
\end{titlepage}

\section{Introduction}

The $q$-state Potts model has served as a valuable model for the study of phase
transitions and critical phenomena \cite{potts,wurev}.  On a lattice, or, more
generally, on a (connected) graph $G$, at temperature $T$, this model is
defined by the partition function
\beq
Z(G,q,v) = \sum_{ \{ \sigma_n \} } e^{-\beta {\cal H}}
\label{zfun}
\eeq
with the (zero-field) Hamiltonian
\beq
{\cal H} = -J \sum_{\langle i j \rangle} \delta_{\sigma_i \sigma_j}
\label{ham}
\eeq
where $\sigma_i=1,...,q$ are the spin variables on each vertex $i \in G$;
$\beta = (k_BT)^{-1}$; and $\langle i j \rangle$ denotes pairs of adjacent
vertices.  The graph $G=G(V,E)$ is defined by its vertex set $V$ and its edge
(bond) 
set $E$; we denote the number of vertices of $G$ as $n=n(G)=|V|$ and the
number of edges of $G$ as $e(G)=|E|$.  We use the notation
\beq
K = \beta J \ , \quad a = e^K = u^{-1} \ , \quad v = a-1
\label{kdef}
\eeq
so that the physical ranges are (i) $a \ge 1$, i.e., $v \ge 0$ corresponding to
$\infty \ge T \ge 0$ for the Potts ferromagnet (FM), with $J > 0$, and (ii) $0
\le a \le 1$, i.e., $-1 \le v \le 0$, corresponding to $0 \le T \le \infty$ for
the Potts antiferromagnet (AFM), with $J < 0$.  One defines the (reduced) free
energy per site $f=-\beta F$, where $F$ is the actual free energy, via
\beq
f(\{G\},q,v) = \lim_{n \to \infty} \ln [ Z(G,q,v)^{1/n}] 
\label{ef}
\eeq
where we use the symbol $\{G\}$ to denote $\lim_{n \to \infty}G$ for a given
family of graphs. 

On a two-dimensional lattice, for the $q=2$ Ising case, and for $q=3,4$, the
Potts ferromagnet exhibits a second-order phase transition from a paramagnetic
(PM) high-temperature phase to a low-temperature phase with spontaneously
broken symmetry and long-range ferromagnetic order (magnetization).  For $q >
4$, this transition is first-order, with a latent heat that increases
monotonically with $q$, approaching a limiting constant as $q \to \infty$
\cite{wurev,baxterbook}. 
The behavior of the Potts antiferromagnet depends on the value of $q$ and the 
type of lattice, as will be discussed further below. 

Let $G^\prime=(V,E^\prime)$ be a spanning subgraph of $G$, i.e. a subgraph
having the same vertex set $V$ and a subset of the edge set, $E^\prime
\subseteq E$. Then $Z(G,q,v)$ can be written as the sum \cite{kf}
\beq
Z(G,q,v) = \sum_{G^\prime \subseteq G} q^{k(G^\prime)}v^{e(G^\prime)}
\label{cluster}
\label{zpol}
\eeq
where $k(G^\prime)$ denotes the number of connected components of $G^\prime$.
Since we only consider connected graphs $G$, we have $k(G)=1$. The formula
(\ref{cluster}) enables one to generalize $q$ from ${\mathbb Z}_+$ to ${\mathbb
R}_+$ (keeping $v$ in its physical range).  The formula (\ref{cluster}) shows
that $Z(G,q,v)$ is a polynomial in $q$ and $v$ (equivalently, $a$).  The Potts
model partition function on a graph $G$ is essentially equivalent to the Tutte
polynomial \cite{tutte1}-\cite{tutte3} and Whitney rank polynomial 
\cite{wurev}, \cite{bbook}-\cite{boll} for this graph, and this connection will
be useful below. 

Using the formula (\ref{cluster}) for $Z(G,q,v)$, one can generalize $q$ from
${\mathbb Z}_+$ not just to ${\mathbb R}_+$ but to ${\mathbb C}$ and $v$ from
its physical ferromagnetic and antiferromagnetic ranges $0 \le v \le \infty$
and $-1 \le v \le 0$ to $v \in {\mathbb C}$.  A subset of the zeros of $Z$ in
the two-complex dimensional space ${\mathbb C}^2$ defined by the pair of
variables $(q,v)$ form an accumulation set in the $n \to \infty$ limit, denoted
${\cal B}$, which is the continuous locus of points where the free energy is
nonanalytic.  The program of studying statistical mechanical models with
external field generalized from ${\mathbb R}$ to ${\mathbb C}$ was pioneered by
Yang and Lee \cite{yl}, and the corresponding generalization of the temperature
from physical to complex values was initiated by Fisher \cite{fisher}.  Here we
allow both $q$ and the temperature-like variable $v$ to be complex.  For a
given value of $v$, one can consider this locus in the $q$ plane, and we shall
sometimes denote it as ${\cal B}_q$, and similarly, for a given value of $q$
(not necessarily $\in {\mathbb Z}_+$), one can consider this locus in the plane
of a complex-temperature variable such as $v$ or
\beq
\zeta = \frac{v}{\sqrt{q}} \ . 
\label{zeta}
\eeq
It will be convenient to introduce polar coordinates, letting
$\zeta=|\zeta|e^{i\theta}$. 

\vspace{4mm}

In this paper we shall present exact calculations of the Potts model partition
function $Z(G,q,v)$ for arbitrary $q$ and $v$ on self-dual strip graphs $G$ of
the square lattice with fixed width $L_y$ and arbitrarily great length $L_x$
with two types of boundary conditions.  Letting $L_x \to \infty$, we compute
the resultant free energy and complex-temperature phase diagram.  Results are
analyzed for widths $L_y=1,2,3$.  We shall use these results to study the
approach to the large-$q$ limit of ${\cal B}$.

There are several motivations for this study.  Clearly, new exact calculations
of Potts model partition functions on lattice strips with arbitrarily large
numbers of vertices are of value in their own right.  This is especially the
case since the free energy of the Potts model has never been calculated exactly
for $d \ge 2$ except in the $q=2$ Ising case in 2D. Just as the study of
functions of a complex variable can yield a deeper understanding of functions
of a real variable, so also the investigation of complex-temperature phase
diagrams of spin models can provide further understanding of the physical
behavior of these models.  Besides \cite{fisher}, complex-temperature
singularities were noticed in early series analyses (e.g., \cite{dombgut}), and
many studies have been carried out on complex-temperature (Fisher) zeros of the
partition function of the Ising model and its generalization to the $q$-state
Potts model \cite{abe}-\cite{css}.  In particular, several exact determinations
of complex-temperature phase diagrams of the Potts model on infinite-length,
finite-width lattice strips \cite{is1d,bcc,a,ta,hca,ka,s3a}, in comparison with
both exact solutions for the $q=2$ 2D complex-temperature phase diagrams
\cite{fisher,chisq,chitri} and finite-lattice calculations of Fisher zeros
\cite{mbook,chw,wuetal,pfef,p,p2} have shown that, although the physical
thermodynamic properties of these strips are essentially one-dimensional, one
can nevertheless gain important insights into certain complex-temperature
properties of the model on the corresponding two-dimensional lattice.  For a
model above its lower critical dimensionality, a complex-temperature phase
diagram includes the complex-temperature extensions of the paramagnetic and
ferromagnetic phases as well as a possible antiferromagnetic phase and other
phases (denoted O in \cite{chisq}) that have no overlap with any physical
phase.  For the infinite-length finite-width strips under consideration here,
for finite $q$, the complex-temperature phase diagram includes only PM and O
phases since there are no broken-symmetry phases.  

An early study of the complex-temperature phase diagram for the square-lattice
Potts model led to the suggestion that the locus ${\cal B}$, which is comprised
of the circles $|a \pm 1|=\sqrt{2}$ for $q=2$ \cite{fisher} might generalize to
the union of the circles $|a-1|=\sqrt{q}$ and $|a+1|=\sqrt{4-q}$ for $1 \le q
\le 4$ \cite{mr}, but subsequent studies found that many of the Fisher zeros in
the $Re(a) < 0$ half-plane do not lie on a circular arc but instead show
considerable scatter \cite{mm,mbook,chw,pfef}.  The infinite square lattice is
self-dual, and for calculations on finite sections of the square lattice, it
was found to be useful to employ boundary conditions that preserve this
self-duality.  Stated more abstractly, one studies lattice graphs $G$ with the
property that the planar dual, $G^*$, is isomorphic to $G$, which we write
simply as $G=G^*$.  Here we recall that the dual of a planar graph $G$ with $n$
vertices, $e$ edges, and $f$ faces is the graph $G^*$ obtained by associating a
vertex of $G^*$ with each face of $G$ and connecting each pair of vertices on
$G^*$ by edges running through the edges of $G$.  It follows that
$n(G^*)=f(G)$, $e(G^*)=e(G)$, and $f(G^*)=n(G)$.  The Potts model partition
function satisfies the relation
\beq
Z(G,q,v)=q^{1-f(G)}v^{e(G)}Z(G^*,q,v_d)
\label{zdual}
\eeq
where the dual image of $v$ is $v_d$, given by
\beq
v_d = \frac{q}{v}
\label{vdual}
\eeq
i.e., in terms of $\zeta$, the duality map is the inversion map:
\beq
\zeta_d = \frac{1}{\zeta} \ .
\label{zetainv}
\eeq
Thus, it is also useful to plot Fisher zeros in terms of the variable
$\zeta$, since the accumulation set ${\cal B}$ is invariant under inversion for
a self-dual graph:
\beq
G = G^* \Longrightarrow {\cal B} \quad {\rm is \ invariant \ under} \quad 
\zeta \to \frac{1}{\zeta} \ .
\label{binv}
\eeq
It was found that complex-temperature zeros calculated for finite sections of
the square with duality-preserving boundary conditions (DBC's) have the
appealing property of lying exactly on an arc of the unit circle $|\zeta|=1$ in
the $\zeta$ plane for $Re(\zeta) \gsim 0$ \cite{mbook,chw,wuetal}. 
(For physical temperature the coefficients of powers of $a=e^K$ are positive,
so there are no zeros on the positive real axis $Re(a) > 0$ for any finite
lattice. However, in the thermodynamic limit, the phase boundary crosses this
axis at $a_c=1+\sqrt{q}$, i.e. $\zeta_c=1$.)  In \cite{pfef} several types of
different self-dual boundary conditions were used in order to ascertain which
features are common to each of these and hence might be relevant for the
thermodynamic limit.  However, because of the scatter of zeros
in the $Re(a) < 0$ half-plane and the dependence of the pattern of zeros, it
has not so far been possible to reach a conclusion concerning the 
portion of the complex-temperature phase boundary ${\cal B}$ in this region 
(although some points on the boundary have been reliably located by analysis 
of series expansions \cite{pfef}). 

An interesting exact result was obtained by Wu and collaborators, who gave an
elegant proof \cite{wuetal}, using Euler's identity for partitions, that for
the Potts model on the square lattice, after having taken the thermodynamic
limit so that zeros of $Z$ in the complex $\zeta$ plane have merged to form the
continuous locus ${\cal B}$, if one takes the further limit $q \to \infty$,
this locus ${\cal B}$ is the circle $|\zeta|=1$ for the square lattice.  It is
straightforward to show that the same conclusion holds if one keeps $L_y$ fixed
and finite, and takes $L_x \to \infty$, after which one takes the limit $q \to
\infty$. Thus, for the infinite-length limit of the self-dual strips that we
consider here, $\lim_{q \to \infty}{\cal B}$ is again the unit circle
$|\zeta|=1$.  In 2D, the interior and exterior of this circle $|\zeta|=1$ form
the complex-temperature extensions of the PM and FM phases, respectively.
Since the infinite-length, finite-width strips under study here are
quasi-one-dimensional systems, there cannot be any broken-symmetry phase at
finite temperature for a spin model with short-range interactions and hence
there is no FM phase or its complex-temperature extension.  Note that any 
finite temperature point, i.e. $-1 < v < \infty$ gets mapped to $\zeta=0$ in
the limit $q \to \infty$.  The identification of phases thus proceeds as
follows in this limit, for the infinite-length finite width strips: the
interior of the circle $|\zeta|=1$ is the PM phase, since it is analytically
connected to the infinite-temperature point $v=\zeta=0$.  The phase in the 
exterior may be interpreted as equivalent to the $T=0$ point, in the sense that
a finite value of $\zeta$ in this region is obtained by taking the double limit
$q \to \infty$ and $|v| \to \infty$ with $v/\sqrt{q}$ held fixed.

Since, as noted above, for finite $q$, the actual pattern of zeros calculated
on finite sections of the square in the $Re(\zeta) < 0$ half-plane show
considerable scatter \cite{mbook,chw,pfef}, two questions arise naturally;
first, having taken the 2D thermodynamic limit, if one starts with $1/q=0$ and
increases this quantity from zero, is there a finite interval in which the
locus ${\cal B}$ continues to be the circle $|\zeta|=1$ before there are
deviations, or do these deviations occur for any finite value of $1/q$, no
matter how small.  We shall address this question here.  A different question
can also be posed for a finite section of the square lattice: for such a
section, with a given size and given duality-preserving boundary conditions, is
there a range in $z=1/q$ above zero in which all of the finitely many Fisher
zeros still occur on the circle $|\zeta|=1$ or not.  This has been considered
in \cite{chw,pfef,kc} and we shall not pursue it here.  Since one does not have
an exact solution for the 2D Potts model for arbitrary $q$, and hence also no
solution for the complex-temperature phase boundary ${\cal B}$, it has not been
possible to determine the precise behavior of this locus analytically in the $q
\to \infty$ limit.

Here one sees the value of exact solutions for the Potts model free energy
and resultant complex-temperature boundary ${\cal B}$ on infinite-length,
finite-width strips, since for these strips, one can obtain exact analytic
answers to the behavior of ${\cal B}$ in the $q \to \infty$ limit.  As we shall
discuss, it is easy to see that one aspect of this behavior is special to the
quasi-one-dimensional nature of the infinite-length strips and is not relevant
for the 2D model, namely that as $1/q$ increases from 0, a gap opens in the
circle $\zeta=1$.  This simply reflects the fact that for a
quasi-one-dimensional spin model with short-range interactions there is no
finite-temperature phase transition and the free energy is analytic for all
finite temperatures, and hence for $0 \le \zeta < \infty$.  However, this is
not a drawback of the method, since finite-lattice calculations of Fisher zeros
on sections of the square lattice have shown that they lie nicely on the circle
$|\zeta|=1$ in the region near the point $\zeta=1$ (while avoiding the precise
point $\zeta=1$ if $n$ is finite).  Hence, one can infer that in the
thermodynamic limit, as $1/q$ is increased from 0, this portion of ${\cal B}$
will remain as an arc of the unit circle.  Indeed, from general arguments
\cite{chisq}, one knows that for the model above its lower critical
dimensionality, where there is a ferromagnetic phase, the portion of the phase
boundary ${\cal B}$ that separates the complex-temperature extension of the
paramagnetic phase from the FM phase must remain intact for finite as well as
infinite $q$.  If it were to bifurcate, this would imply a new third phase
between the PM and FM phases, contrary to the known properties of the Potts
model, so, given the invariance of ${\cal B}$ under the inversion symmetry
(\ref{binv}), this portion must remain on the circle $|\zeta|=1$.

Since the explicit calculations of Fisher zeros showed large scatter away from
$|\zeta|=1$ in the $Re(\zeta) < 0$ half-plane even for values well above $q=4$,
such as $q=10$ \cite{wuetal,pfef}, this suggested that the totality of Fisher
zeros would only cluster on the circle $|\zeta|=1$ in the limit $q \to \infty$
itself but some zeros, and some portion of their accumulation set ${\cal B}$
would deviate from it for all finite $q$.  Calculations using the usual
Hamiltonian formulation of the Potts model and associated transfer matrix
methods become increasingly cumbersome for large $q$ because of the
increasingly many states.  For the purpose of studying the large-$q$ behavior
of the zeros, it is convenient to solve for $Z(G,q,v)$ for arbitrary $q$ and
$v$ on large finite lattice sections.  This was done in \cite{ks} and the zeros
were calculated for $q$ up to 100; again these showed only a slow approach to
the circle $|\zeta|=1$.  Together with the previous calculations for $q$ up to
10, it was concluded in \cite{ks} that this evidence supported the inference
that the totality of Fisher zeros only lie on the circle $|\zeta|=1$ in the
limit $q \to \infty$.  This type of calculations has also been done in
\cite{kc} with the same conclusion.  Our exact results on infinite-length
finite-width strips complement these finite-lattice calculations and allow a
rigorous conclusion that for these strips ${\cal B}$ deviates from $|\zeta|=1$
for any $1/q$ no matter how small.

These results are relevant in another way.  Large-$q$ series expansions (in the
variable $1/\sqrt{q}$) have been useful in studying the thermodynamic
properties of Potts models \cite{lacaze}-\cite{arisue99b}.  Large-$q$
expansions (in the variable $1/(q-1)$) have also been useful for a particular
special case of the Potts model, namely the $T=0$ special case of the Potts
antiferromagnet, where the partition function on a graph $G$ reduces to the
chromatic polynomial,
\beq 
Z(G,q,-1)=P(G,q)
\label{zp}
\eeq
where $P(G,q)$ is the chromatic polynomial (in $q$) expressing the number of
ways of coloring the vertices of the graph $G$ with $q$ colors such that no two
adjacent vertices have the same color \cite{bbook,rtrev}.  Indeed, for a
given graph $G$ and for sufficiently large $q$, the Potts antiferromagnet
exhibits nonzero ground state entropy (without frustration).  This is
equivalent to a ground state degeneracy per site (vertex), $W > 1$, since $S_0
= k_B \ln W$, where $W(\{G\},q)= \lim_{n \to \infty}P(G,q)^{1/n}$.  Large-$q$
series expansions for $W_r(\{G\})=q^{-1}W(\{G\},q)$ have been given, e.g., in
\cite{nagle}-\cite{wn}. 

A different motivation for the present study is the following.  Just as was
true for the double-complexification of field and temperature studied in
\cite{ih}-\cite{yy}, where one gained a deeper understanding of the singular
locus (continuous accumulation set of partition function zeros) by considering
the separate projections in the planes of complex-field and complex temperature
by considering these as parts of a single underlying submanifold (with possible
singularities) in the ${\mathbb C}^2$ space of complex temperature and field,
so also here, one gains similar insight into the projections of ${\cal B}$ in
the complex $q$ and $v$ plane by relating these as different slices of the
locus ${\cal B}$ in the ${\mathbb C}^2$ space defined by $(q,v)$.

\section{Generalities}

We refer the reader to our earlier papers containing exact calculations of
$Z(G,q,v)$ for a number of further details.  We recall that 
the formal definition of the free energy may be insufficient to define this
function at certain special special values $q=q_s$ \cite{a}; it is
necessary to specify the order of the limits that one uses. 
We denote the limits with the two different orders as 
definitions using different orders of limits as $f_{qn}$ and $f_{nq}$:
\beq
f_{nq}(\{G\},q,v) = \lim_{n \to \infty} \lim_{q \to q_s} n^{-1} \ln Z(G,q,v)
\label{fnq}
\eeq
\beq
f_{qn}(\{G\},q,v) = \lim_{q \to q_s} \lim_{n \to \infty} n^{-1} \ln Z(G,q,v) \
.
\label{fqn}
\eeq
Where necessary, we shall comment on which order of limits is used below. 
Of course in discussions of the usual $q$-state Potts model (with
positive integer $q$), one automatically uses the definition in eq.
(\ref{zfun}) with (\ref{ham}) and no issue of orders of limits arises, as it
does in the Potts model with real $q$.
As a consequence of the above noncommutativity, it follows that for
the special set of points $q=q_s$ one must distinguish between (i) $({\cal
B}_a(\{G\},q_s))_{nq}$, the continuous accumulation set of the zeros of
$Z(G,q,v)$ obtained by first setting $q=q_s$ and then taking $n \to \infty$,
and (ii) $({\cal B}_a(\{G\},q_s))_{qn}$, the continuous accumulation set of the
zeros of $Z(G,q,v)$ obtained by first taking $n \to \infty$, and then taking $q
\to q_s$.  For these special points,
\beq
{\cal B}_{nq} \ne {\cal B}_{qn} \ .
\label{bnoncom}
\eeq
We have discussed this type of noncommutativity in earlier papers (e.g.,
\cite{w,a}). 

A general form for the Potts model partition function
for the strip graphs considered here, or more generally, for recursively
defined families of graphs comprised of $m$ repeated subunits (e.g. the columns
of squares of height $L_y$ vertices that are repeated $L_x$ times to form an
$L_x \times L_y$ strip of a regular lattice with some specified boundary
conditions), is \cite{a} 
\beq 
Z(G,q,v) = \sum_{j=1}^{N_{Z,G,\lambda}} c_{G,j}(\lambda_{G,j}(q,v))^m
\label{zgsum}
\eeq
where the terms $\lambda_{G,j}$, the coefficients $c_{G,j}$, and the total
number $N_{Z,G,\lambda}$ depend on $G$ through the type of lattice, its width,
$L_y$, and the boundary conditions, but not on the length.  Following our
earlier nomenclature \cite{w}, we denote a $\lambda$ as leading (= dominant) if
it has a magnitude greater than or equal to the magnitude of other $\lambda$'s.
In the limit $n \to \infty$ the leading $\lambda$ in $Z$ determines the free
energy per site $f$.  The continuous locus ${\cal B}$ where $f$ is nonanalytic
thus occurs where there is a switching of dominant $\lambda$'s in $Z$ and $P$,
respectively, and is the solution of the equation of degeneracy in magnitude of
these dominant $\lambda$'s.  The special case of this for the chromatic
polynomial was discussed in \cite{bkw,bkw80}.

Let us next define the self-dual strip graphs that we consider here.  The first
type, with boundary conditions that we denote as DBC1 (following our
nomenclature in \cite{pfef}) was discussed in Ref. \cite{chw} and illustrated
in Fig. 1 of that paper.  We shall need a straightforward generalization of it
to the case of an $L_x \times L_y$ lattice with $L_x \ne L_y$, and we describe
this as follows.  Let the lattice be oriented with the $x$ and $y$ directions
being horizontal and vertical, respectively.  Let all of the vertices on the
upper and right-hand sides, including the corner vertices, connect along
directions outward from the lattice to a common vertex adjoined to this lattice
(so that the upper right corner connects to this adjoined vertex via edges in
both the $x$ and the $y$ directions), while all of the sites on the lower and
left-hand edges, excluding the previously mentioned corner vertices, have free
boundary conditions.  It is easily checked that this graph is self-dual.  Note
that it has free longitudinal (horizontal) boundary conditions.  The number of
vertices is $n=L_xL_y+1$.  Graphs of this type were used for calculations of
Fisher zeros in \cite{chw,pfef}.  

\begin{figure}[h]
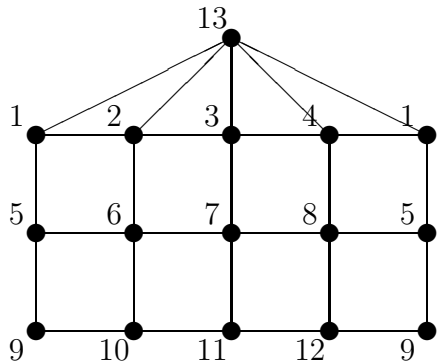

\vspace{1cm}
\caption{\footnotesize{Example of an $L_y \times L_x$ strip of the square
lattice with DBC2 boundary conditions, for the case $L_y=3$, $L_x=4$.}}
\label{strip}
\end{figure}

The second type of self-dual strip graph, used in \cite{pfef} where it was
labelled DBC2 \cite{wup}, can be described as follows.  Let the $L_x \times
L_y$ lattice strip have periodic boundary longitudinal (=horizontal) boundary
conditions and connect all of the vertices on the upper side of the strip to a
single external vertex, while all of the vertices on the lower side of the
strip have free boundary conditions \cite{wup}.  An illustration of this type
of graph is given in Fig. \ref{strip}.  This has also recently been used for
calculation of Fisher zeros in \cite{kc}.  In \cite{dg} we gave exact results
for structural properties of Potts model partition functions and chromatic
polynomials for strips of this type, of arbitrarily great length and width, and
presented exact calculations of chromatic polynomials and resultant singular 
loci ${\cal B}$ for $v=-1$ in the $q$ plane for widths up to $L_y=4$. 

\vspace{8mm}

\unitlength 1.3mm
\begin{picture}(40,30)
\multiput(30,0)(10,0){5}{\circle*{2}}
\multiput(30,10)(10,0){5}{\circle*{2}}
\multiput(30,20)(10,0){5}{\circle*{2}}
\multiput(30,0)(10,0){5}{\line(0,1){20}}
\multiput(30,0)(0,10){3}{\line(1,0){40}}
\put(50,30){\circle*{2}}
\put(30,20){\line(2,1){20}}
\put(40,20){\line(1,1){10}}
\put(50,20){\line(0,1){10}}
\put(60,20){\line(-1,1){10}}
\put(70,20){\line(-2,1){20}}
\put(28,-2){\makebox(0,0){9}}
\put(38,-2){\makebox(0,0){10}}
\put(48,-2){\makebox(0,0){11}}
\put(58,-2){\makebox(0,0){12}}
\put(68,-2){\makebox(0,0){9}}
\put(28,12){\makebox(0,0){5}}
\put(38,12){\makebox(0,0){6}}
\put(48,12){\makebox(0,0){7}}
\put(58,12){\makebox(0,0){8}}
\put(68,12){\makebox(0,0){5}}
\put(28,22){\makebox(0,0){1}}
\put(38,22){\makebox(0,0){2}}
\put(48,22){\makebox(0,0){3}}
\put(58,22){\makebox(0,0){4}}
\put(68,22){\makebox(0,0){1}}
\put(48,32){\makebox(0,0){13}}
\end{picture}

We comment on an interesting feature of the complex-temperature phase diagrams
for self-dual infinite-length, finite-width strips.  In our earlier work
yielding exact determinations of these complex-temperature phase diagrams for
non-self-dual strips \cite{a,bcc,ta,hca,ka,s3a}, it was found that for some
cases, e.g. strips with periodic longitudinal boundary conditions, ${\cal B}$
passes through the origin of the $u=a^{-1}$ plane.  For free longitudinal
boundary conditions, this does not happen.  For the self-dual strip graphs
considered here, we can easily prove that ${\cal B}$ does not pass through
$u=0$, i.e $v=\zeta=\infty$, since by the inversion symmetry under $\zeta \to
1/\zeta$, this would imply that it passes through $\zeta=0$, but this is the
infinite-temperature ($\beta=0$) point, where the free energy is analytic, so
no singular phase boundary can pass through this point.

\section{$L_{\lowercase{y}}=1$ Strip with DBC1}

In this section we present the Potts model partition function $Z(L_y \times
L_x,DBC1,q,v)$ for the strips of the square lattice of width $L_y$ and
arbitrarily great length $L_x=m+1$ containing $m$ edges in each horizontal row
of the strip, with duality-preserving boundary conditions of type 1.  We label
such a strip graph with DBC1 boundary conditions as $S,L_y$ or just $SL_y$ for
short and $(S,L_y)_m$ to indicate the length.  The
number of vertices is $n=L_xL_y+1$.  One convenient way to express the results
is in terms of a generating function,
\beq \Gamma(S,L_y,q,v,z)= \sum_{m=0}^\infty Z(S,L_y,m,q,v)z^m \ . 
\label{gammazfbc}
\eeq
As indicated, the coefficients in the Taylor series expansion of this
generating function in the auxiliary variable $z$ are the partition functions
for the strip of length $m$.  We have calculated this generating function using
transfer matrix methods and iterative application of the deletion-contraction
theorem for the corresponding Tutte polynomial.  We find
\beq \Gamma(S,L_y,q,v,z) = \frac{ {\cal N}(S,L_y,q,v,z)}
{{\cal D}(S,L_y,q,v,z)}
\label{gammazcalc}
\eeq
where the numerator ${\cal N}(S,L_y,q,v,z)$ and the denominator
${\cal D}(S,L_y,q,v,z)$ are polynomials in $z$, $q$, and $v$ that depend on
$L_y$ but not $L_x$. The degree of the denominator in $z$, i.e., the number of
$\lambda$'s in the form (\ref{zgsum}), is \cite{dg} 
\beq
N_{Z,S,L_y,\lambda} = deg_z({\cal D}(S,L_y,q,v,z))=
\frac{2}{L_y+2}{2L_y+1 \choose L_y} \ . 
\label{nztotdbc1}
\eeq

We first treat the minimum-width case, $L_y=1$, which has the appeal that the
analytic results are simple but already exhibit a rich variety of behavior for
the locus ${\cal B}$.  This family of graphs can be represented as an open
wheel formed by $m+1$ vertices along the rim, each except the rightmost one
connected by a spoke (edge) with a vertex forming the axle of the open wheel,
and with the rightmost vertex on the rim connected by a double edge to this
central vertex.  We calculate for the denominator of the generating function
(with the abbreviation $S1$ for $S,L_y=1$)
\beqs
{\cal D}(S1,q,v,z) & = & 1-(3v+q+v^2)z+v(v+1)(v+q)z^2 \cr\cr
& = & \prod_{j=1}^2 (1-\lambda_{S1,j}z)
\label{ds1}
\eeqs
where
\beq
\lambda_{S,1,(1,2)}=\frac{1}{2}\biggl [ T_{S1} \pm \sqrt{R_{S1}} \ \biggr ]
\label{lams112}
\eeq
with 
\beqs
T_{S1} & = & 3v+q+v^2 \cr\cr
       & = & 3\sqrt{q}\zeta+q+q\zeta^2
\label{ts1}
\eeqs
\beqs
R_{S1} & = & 5v^2+2vq+2v^3+q^2-2v^2q+v^4 \cr\cr
       & = & 
5q\zeta^2+2q^{3/2}\zeta+2q^{3/2}\zeta^3+q^2-2q^2\zeta^2+q^2\zeta^4
\label{rs1}
\eeqs
and for the numerator of the generating function 
\beq
{\cal N}(S,1,q,v,z)=A_{S1,0}+A_{S1,1}z
\label{numgamma}
\eeq
with
\beq
A_{S1,0}=q(2v+q+v^2)
\label{as10}
\eeq
\beq
A_{S1,1}=-qv(v+1)(v+q) \ . 
\label{as11}
\eeq
Ref. \cite{hs} presented a formula to obtain the chromatic polynomial
for a recursive family of graphs in the form of sums of powers of
$\lambda_j$'s starting from the generating function, and the
generalization of this to the full Potts model partition function was
given in \cite{a}.  Using this, we have 
\beq
Z(S1_m,q,v) = \frac{(A_{S1,0}\lambda_{S1,1} + A_{S1,1})}
{(\lambda_{S1,1}-\lambda_{S1,2})}(\lambda_{S1,1})^m +
 \frac{(A_{S1,0} \lambda_{S1,2} + A_{S1,1})}
{(\lambda_{S1,2}-\lambda_{S1,1})}(\lambda_{S1,2})^m \ . 
\label{pgsumkmax2}
\eeq 
It is readily verified that this is symmetric under the interchange
$\lambda_{S1,1} \leftrightarrow \lambda_{S1,2}$.  The free energy is given by
$f=\ln \lambda_{S1,1}$ and is analytic for all finite temperature.  Regarding
the locus ${\cal B}$ as a submanifold (with possible singularities) in the
${\mathbb C}^2$ space defined by the variables $(q,v)$ or $(q,\zeta)$, we can
obtain the slices of this locus in the complex $q$ plane for fixed $v$ and in
the complex $v$ or $\zeta$ plane for fixed $q$.

\subsubsection{$L_y=1$ with DBC1: ${\cal B}_q$ for fixed $v$}

For the physical range $v \in [-1, \infty]$, the locus ${\cal B}$
in the $q$ plane consists of a single self-conjugate arc which has endpoints at
\beq
q_e, q_e^* = v(v-1 \pm 2i\sqrt{v+1} \ )
\label{qes1}
\eeq
and crosses the real $q$ axis at 
\beq
q=-v(v+3)  \ .
\label{qcross_s1}
\eeq
These points $q_e,q_e^*$ are the branch points of the square root
$\sqrt{R_{S1}}$. For the Potts antiferromagnet at $T=0$, i.e., $v=-1$, the
locus ${\cal B}$ degenerates to the single point $q=2$.  As one increases the
temperature above $T=0$, ${\cal B}$ expands to form the generic arc as given 
above, but then as $T \to \infty$, i.e., $v \to 0^-$, this arc shrinks again to
a point at the origin, $q=0$.  For the Potts ferromagnet, as $T$ decreases from
infinity, i.e., $v$ increases from 0, the arc is centered in the negative
$Re(q)$ half-plane and crosses the negative real $q$ axis at the value given in
eq. (\ref{qcross_s1}).

\subsubsection{$L_y=1$ with DBC1: ${\cal B}_{\zeta}$ for fixed $q$}

We discuss here the locus ${\cal B}$ in the complex $\zeta$ plane for fixed
$q$.  We shall concentrate on the range of real $q \ge 1$.  The locus ${\cal
B}$ is defined by the equality of magnitudes $|\lambda_{S,1,1}|=
|\lambda_{S,1,2}|$.  This equality can arise in two, in general separate, ways.
First, (for real $q$), on the real axis of the $\zeta$ plane, since $T_{S1}$ is
real, if $R_{S1} < 0$ so that the square root in (\ref{lams112}) is pure
imaginary, it follows that $|\lambda_{S,1,1}|=|\lambda_{S,1,2}|$.  Second, at
the four points where $R_{S1}=0$, clearly $\lambda_{S,1,1}=\lambda_{S,1,2}$.
At certain special values of $q$ some of these six points can coincide.

\begin{figure}[hbtp]
\vspace{-10mm}
\centering
\leavevmode
\epsfxsize=3in
\begin{center}
\leavevmode
\epsffile{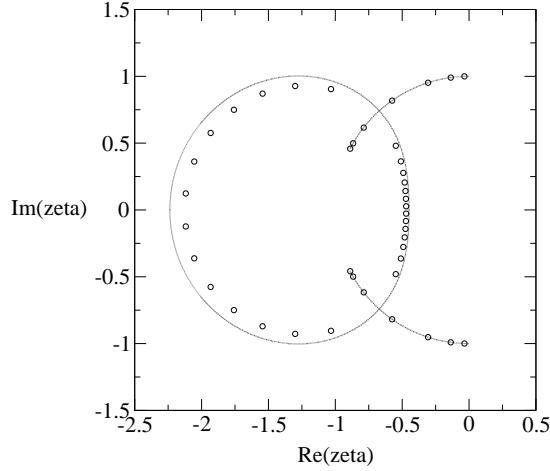}
\end{center}
\vspace{-10mm}
\caption{\footnotesize{Singular locus ${\cal B}$ in the $\zeta$
plane for the free energy of the $q=5/4$ Potts model on
the $L_x \to \infty$ limit of the $L_y=1$ strip with DBC1 boundary
conditions. For comparison, zeros of $Z$ for $L_x=21$ are shown.}}
\label{why1vq1p25}
\end{figure}

Proceeding to analyze ${\cal B}$, we first observe that at $q=1$, this locus is
comprised of a closed oval curve that surrounds the point $\zeta=-1$ and
crosses the real $\zeta$ axis at $\zeta=(1/2)(-3 \pm \sqrt{5} \ )$, i.e., at
$\zeta \simeq -2.618$ and $-0.3820$. The crossing point on the left is at 
$\zeta=-B_5$, where
\beq
B_r = 4\cos^2\bigg ( \frac{\pi}{r} \bigg )
\label{br}
\eeq
is the Tutte-Beraha number.  Recall here that ${\cal B} \equiv {\cal B}_{qn}$;
if one were to use the opposite order of limits in (\ref{bnoncom}), then
$Z(G,q=1,v)=(v+1)^n$ and all zeros collapse to the single point $v=\zeta=-1$.
This illustrates the noncommutativity discussed in the introduction.  As $q$
increases above $q=1$, two complex-conjugate arcs sprout out from the points
$\zeta=e^{\pm 2 \pi i/3}$, so that ${\cal B}$ is comprised of the union of
these arcs and the closed oval curve. The endpoints of the arcs occur at the
four zeros of $R_{S1}$, i.e., branch point zeros of $\sqrt{R_{S1}}$.  The
right-hand endpoints of the arc are located at the complex conjugate pair of
points
\beq
\zeta_{ae},\zeta_{ae}^* = q^{-1/2}\bigg [ \sqrt{q-1} - \frac{1}{2} 
\pm i\bigg (\sqrt{q-1} + \frac{3}{4} \bigg )^{1/2} \bigg ]
\label{zetaae}
\eeq
at the angles $\pm \theta_{ae}$ given by 
\beq
\theta_{ae} = {\rm arctan} \Bigg [ \frac{\bigg [3+4\sqrt{q-1} \bigg ]^{1/2}}
{2\sqrt{q-1}-1} \Bigg ] \ . 
\label{theta_ae}
\eeq
The left-hand endpoints occur at 
\beq
\zeta_{se},\zeta_{se}^{-1} = q^{-1/2}\bigg [ -\sqrt{q-1} - \frac{1}{2}
\pm \bigg (\sqrt{q-1} - \frac{3}{4} \bigg )^{1/2} \bigg ] \ .
\label{zetase}
\eeq
For $q < q_{ac}$, where 
\beq
q_{ac} = (5/4)^2=1.5625
\label{qac}
\eeq
$\zeta_{se}$ is complex and $\zeta_{se}^{-1}=\zeta_{se}^*$, while for $q \ge
(5/4)^2$, $\zeta_{se}$ is real. For $q=(5/4)^2$, we have $\zeta_{se}=-1$.
Thus, as $q$ increases through the value $q=q_{ac}$, the left-hand endpoints of
the complex-conjugate arcs come together and pinch the negative real axis at
$\zeta=-1$.  For $q > q_{ac}$, this part of the locus ${\cal B}$ forms a line
segment on the negative real axis centered at $\zeta=-1$, whose right-hand end
is $\zeta_{se}$ and left-hand end its inverse.  The above-mentioned oval curve
crosses the negative real $\zeta$ axis at the two points where $T_{S1}=0$,
i.e.,
\beq
\zeta_t,\zeta_t^{-1} = \frac{1}{2\sqrt{q}}\bigg [ -3 \pm \sqrt{9-4q} \ \bigg ]
\ . 
\label{zeta_ts1eq0}
\eeq
For $q=1$, $\zeta_t=(1/2)(-3+\sqrt{5} \ ) \simeq -0.3820$,
$\zeta_t^{-1}=-(1/2)(3+\sqrt{5} \ ) = -B_5$, as discussed above.  The locus
${\cal B}$ and corresponding complex-temperature phase diagram is plotted for a
typical value in the interval $1 < q < q_{ac}$, namely, $q=5/4=1.25$, in
Fig. \ref{why1vq1p25}.  Here $\theta_{ae}=\pi/2$.  For comparison,
complex-temperature Fisher zeros are shown for a long finite strip.  As is
evident, the density decreases strongly as one approaches the intersection
points (multiple points in the terminology of algebraic geometry) where the
arcs and the oval curves cross each other.  This is the same behavior that we
found in many previous studies of complex-temperature zeros for spin models,
e.g., \cite{hs,is1d,only,1dnnn}.  The complex-temperature extension of the
paramagnetic (PM) phase occupies the full $\zeta$ plane except for an O region
enclosed by the oval curve (and the singular set of measure zero comprised by
${\cal B}$ itself).

\begin{figure}[hbtp]
\vspace{-10mm}
\centering
\leavevmode
`\epsfxsize=3in
\begin{center}
\leavevmode
\epsffile{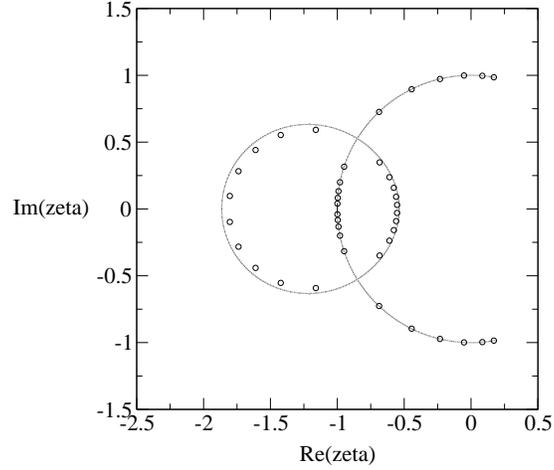}
\end{center}
\vspace{-10mm}
\caption{\footnotesize{Same as Fig. \ref{why1vq1p25} for 
$q=(5/4)^2=1.5625$.}}
\label{why1vq25ov16}
\end{figure}

\begin{figure}[hbtp]
\vspace{-10mm}
\centering
\leavevmode
\epsfxsize=3in
\begin{center}
\leavevmode
\epsffile{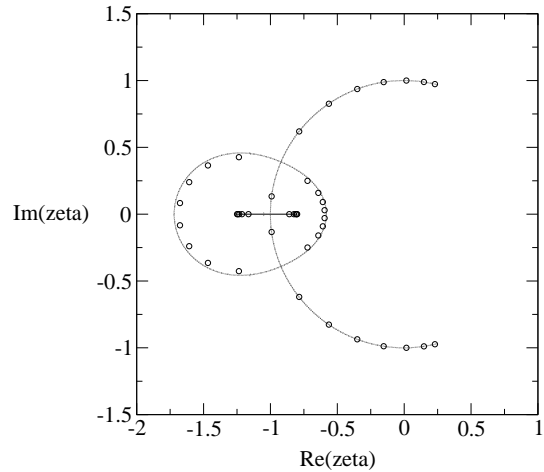}
\end{center}
\vspace{-10mm}
\caption{\footnotesize{Same as Fig. \ref{why1vq1p25} for $q=1.7$.}}
\label{why1vq1p7}
\end{figure}

For an interval $q \ge q_{ac}$, there are two O phases, namely the regions
surrounded by the oval curve and separated by the arc of the $|\zeta|=1$ circle
that passes through $\zeta=-1$.  The rest of the $\zeta$ plane is occupied by
the complex-temperature extension of the paramagnetic (PM) phase.  This type of
situation is illustrated in Fig. \ref{why1vq25ov16} for $q=(5/4)^2$ and 
Fig. \ref{why1vq1p7} for $q=1.7$. 

\begin{figure}[hbtp]
\vspace{-10mm}
\centering
\leavevmode
\epsfxsize=3in
\begin{center}
\leavevmode
\epsffile{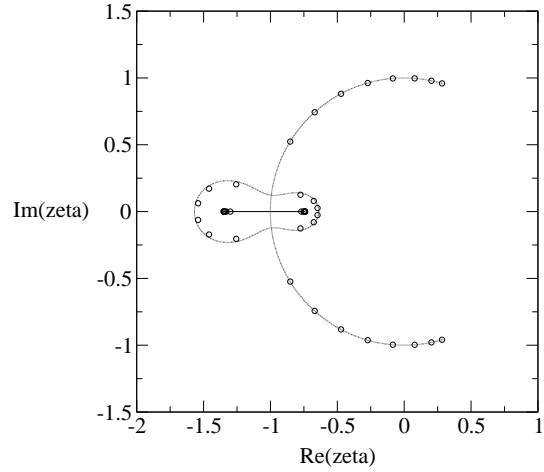}
\end{center}
\vspace{-10mm}
\caption{\footnotesize{Same as Fig. \ref{why1vq1p25} for $q=1.85$.}}
\label{why1vq1p85}
\end{figure}

\begin{figure}[hbtp]
\vspace{-10mm}
\centering
\leavevmode
\epsfxsize=3in
\begin{center}
\leavevmode
\epsffile{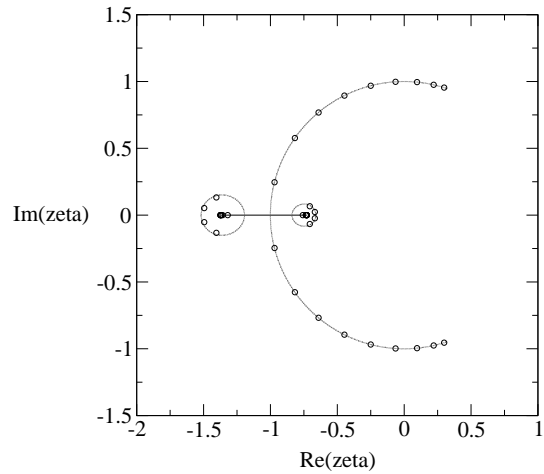}
\end{center}
\vspace{-10mm}
\caption{\footnotesize{Same as Fig. \ref{why1vq1p25} for $q=1.9$.}}
\label{why1vq1p9}
\end{figure}

As $q$ increases further, the two O phases that were contiguous now contract
and pull away from each other.  This is illustrated in Fig.  \ref{why1vq1p85}
for $q=1.85$.  Eventually, these two O phases pull completely away so that they
are no longer contiguous; they are then centered around the endpoints of the
line segment.  An example is shown in Fig. \ref{why1vq1p9} for $q=1.9$.  Here
the crossing of the right-hand O phase, i.e. the crossing nearest to the
origin, occurs at $\zeta_t$ and its inverse is the leftmost crossing of the
other O phase.

\begin{figure}[hbtp]
\vspace{-10mm}
\centering
\leavevmode
\epsfxsize=3in
\begin{center}
\leavevmode
\epsffile{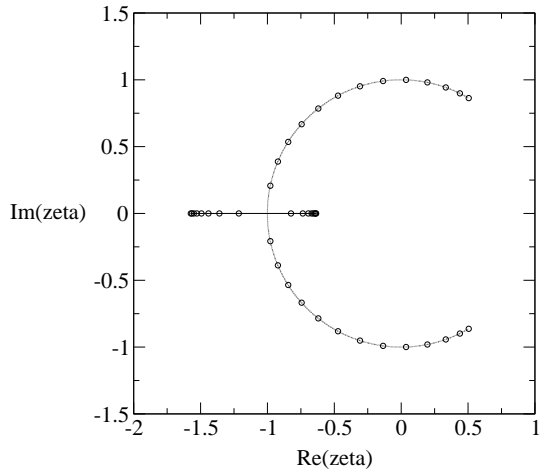}
\end{center}
\vspace{-10mm}
\caption{\footnotesize{Same as Fig. \ref{why1vq1p25} for $q=3$.}}
\label{why1vq3}
\end{figure}

As $q$ increases to $q=2$, the two O phases contract to points and disappear.
One can see this analytically since at $q=2$ the crossing $\zeta_t=-1/\sqrt{2}$
coincides with the line segment endpoint $\zeta_{se}=-1/\sqrt{2}$ and similarly
for their inverses.  In the interval $2 < q < 9/4 = 2.25$ the points $\zeta_t$
and $\zeta_t^{-1}$ are located in the interior of the line segment and do not
play a special role.  As $q$ increases above $q=9/4$, there ceases to be any
real-$q$ solution of the condition $T_{S1}=0$.  The locus ${\cal B}$ and
complex-temperature phase diagram are shown for $q=3$ in Fig. \ref{why1vq3}.  A
topological feature of ${\cal B}$ for this region of $q$, namely the presence
of a complex-temperature endpoint of a line segment on the left, is reminiscent
of the suggestive possibility of prongs (or perhaps cusps) on ${\cal B}$ for
the square-lattice Potts model inferred from the combination of calculations of
Fisher zeros and the correlation of the positions of these zeros with locations
of complex-temperature singularities that were reliably determined from
analyses of low-temperature series expansions in \cite{pfef} (see also
\cite{p,p2}).  The angle $\theta_{ae}$ of the upper arc endpoint decreases as
$q$ increases, i.e., this endpoint moves toward the point $\zeta=1$ on the real
axis.  In Table \ref{table_theta_ae} we list some explicit values of this angle
$\theta_{ae}$ as a function of $q$ for (the $L_x \to \infty$ limit of) this
strip.

\begin{table}
\caption{\footnotesize{Values of arc endpoint angles $\theta_{ae}$ for various
infinite-length finite-width square lattice strips with duality-preserving
boundary conditions. The $\theta_{ae}$ values are the same for DBC1 and DBC2.}}
\begin{center}
\begin{tabular}{|c|c|c|c|c|c|}\hline
$L_y$ & $q=1$ & $q=2$ & $q=3$ & $q=4$ & $q=10$ \\ \hline 
1 & $120^\circ$  & $69.3^\circ$ & $58.1^\circ$ & $52.0^\circ$ & $37.8^\circ$ 
\\ \hline
2 & $88.8^\circ$ & $47.7^\circ$ & $38.8^\circ$ & $34.1^\circ$ & $23.7^\circ$ 
\\ \hline
\end{tabular}
\end{center}
\label{table_theta_ae}
\end{table}

As $q$ gets large, the right-hand arc endpoints move down toward the point
$\zeta=1$ and the circular arc finally pinches this point in the limit as $q
\to \infty$.  We calculate the following expansions for the position of
the upper arc endpoint for large $q$: 
\beq
\zeta_{ae} = 1 +iq^{-1/4} - \frac{1}{2}q^{-1/2} + \frac{3}{2^3}i q^{-3/4} 
- \frac{1}{2}q^{-1} -\frac{41}{2^7}i q^{-5/4} + \frac{123}{2^{10}}i q^{-7/4}
- \frac{1}{2^3} q^{-2} + O(q^{-9/4})
\label{zeta_ae_qinf}
\eeq
(the lower one being the complex conjugate) and, for the right and left 
endpoints of the line segment.  On the left, as $q$ gets large the line segment
contracts toward $\zeta=-1$ and finally degenerates to a point at
$q=\infty$. We calculate the following expansion for the positions of the
endpoints of this line segment for large $q$:
\beqs
& & \zeta_{se}, \zeta_{se}^{-1} = -1 \pm q^{-1/4} - \frac{1}{2}q^{-1/2} 
\mp \frac{3}{2^3}q^{-3/4} + \frac{1}{2}q^{-1} \mp \frac{41}{2^7}q^{-5/4} \cr\cr
& & \mp \frac{123}{2^{10}}q^{-7/4} + \frac{1}{2^3}q^{-2} + O(q^{-9/4}) \ . 
\label{zetaereell}
\eeqs
Thus, in the limit as $q \to \infty$, ${\cal B}$ becomes the unit circle
$|\zeta|=1$.  

Going the other way, let us start at $q=\infty$, where ${\cal B}$ is the circle
$|\zeta|=1$.  As $q$ decreases from infinity, two changes occur immediately in
${\cal B}$: (i) the circle breaks open on the right side, forming two arcs with
endpoints at the angles given in (\ref{theta_ae}) that recede away from the
real axis, and (ii) a real line segment sprouts out from the point $\zeta=-1$.
Feature (i) reflects the quasi-one-dimensional nature of the $L_x \to \infty$
limit of this family of strip graphs, since for finite $q$, the free energy of
the Potts ferromagnet is analytic for all finite temperatures.  This feature
does not hold in the thermodynamic limit on the square lattice.  As was
discussed in \cite{chisq,cmo} for the Ising model and in \cite{pfef} for the
general Potts model, the region around $a=1$, i.e., $v=0$, which is the
paramagnetic phase, is not analytically connected to the broken-symmetry,
ferromagnetic phase; hence the part of the phase boundary ${\cal B}$ that
separates the complex-temperature extensions of the PM phase and FM phases from
each other must remain intact as $q$ decreases from infinity. However, the
deviation on the left serves as a prototype of the sort of deviations that are
suggested by finite-lattice calculations of Fisher zeros for sections of the
square lattice \cite{mbook,wuetal,pfef,kc} and gives some insight, as an
exactly calculable example, of how these deviations arise.  As $q$ decreases to
sufficiently small values, the line segment changes to a region around
$\zeta=-1$.  As we shall show below, the nature of the deviation on the left
can be more complicated for wider strips. Thus, from the point of view of
increasing $q$, ${\cal B}$ only becomes the unit circle $|\zeta|=1$ in the
limit $q \to \infty$.

We next point out an another important feature of these results.  For the $q=3$
case, ${\cal B}$ crosses the real axis at $\zeta=-1/\sqrt{3}$, its inverse
$-\sqrt{3}$, and at $\zeta=-1$.  Transforming back to the $v$ plane, these
points correspond to $v=-1$, $v=-3$, and $v=-\sqrt{3}$, respectively.  The
crossing at $v=-1$, i.e., $a=0$, connotes a zero-temperature critical point of
the $q=3$ Potts antiferromagnet on this infinite-length $L_y=1$ strip
\cite{lenard,lieb}.  This is very interesting, since this model also has a
zero-temperature critical point on the (infinite) square lattice.  Thus, an
infinite-length strip with width $L_y=1$ already exhibits a feature of the
Potts antiferromagnet on the full square lattice.  As will be seen below, this
is also true of the other widths $L_y=2,3$ for which we have obtained exact
solutions for the Potts model free energy.  The crossing at $v=-3$, i.e.,
$a=-2$, is a complex-temperature singular point that is the dual image of $a=0$
and, by duality, the singularity in the free energy is the same as at the
physical zero-temperature critical point, in accordance with the general
discussion in \cite{hcl} relating physical and complex-temperature
singularities by duality.  It is instructive to view the locus ${\cal B}$ in
the complex $\zeta$ or $v$ plane for fixed $q$ as a slice of the singular
subset in the full ${\mathbb C}^2$ space defined by the pair of variables or
$(q,v)$.  Thus, the crossing of ${\cal B}$ at $v=-1$ for $q=3$ is the point
$(q,v)=(3,-1)$ and corresponds to the crossing of the slice of ${\cal B}$ at
$q=3$ in the $q$ plane at fixed $v=-1$.  This is precisely the $q_c$ that we
found previously in our study of chromatic polynomials and their asymptotic
limits and loci ${\cal B}$ for this family of graphs in \cite{dg}.  It will be
recalled that we found that this $q_c=3$ value was universal for all of the
widths $1 \le L_y \le 4$ for which we calculated exact solutions for the
chromatic polynomial and resultant ${\cal B}$.  As we shall show below, this
corresponds to the feature that for each of the widths of strips that we study
here, ${\cal B}$ passes through $v=-1$, i.e., $\zeta=-1/\sqrt{3}$ (and, by
duality, its inverse, $\zeta=-\sqrt{3}$) for $q=3$.

For the Ising case $q=2$, ${\cal B}$ crosses the $v$ axis at $v=-1$ and the
dual image $v=-2$.  The crossing at $v=-1$ connotes a zero-temperature critical
point for the Ising antiferromagnet on the $L_x \to \infty$ limit of this
graph. 

\begin{figure}[hbtp]
\vspace{-10mm}
\centering
\leavevmode
\epsfxsize=3in
\begin{center}
\leavevmode
\epsffile{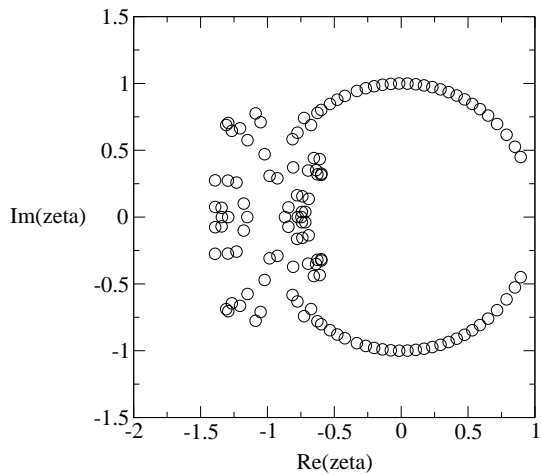}
\end{center}
\vspace{-10mm}
\caption{\footnotesize{Complex-temperature Fisher zeros in the $\zeta$
plane for the $q=4$ Potts model on the $L_y=3$ strip with DBC1 boundary
conditions with $L_x=21$ ($n=64$).}}
\label{why3vq4}
\end{figure}

We have also calculated Fisher zeros for the strips with DBC1 and $L_y=2,3$.  A
typical example is $L_y=3$, $q=4$, shown in Fig. \ref{why3vq4}.  For lack of
space we do not show the others here, but they are available upon request.

\section{$L_{\lowercase{y}}=1$, DBC2}

To elucidate the dependence of ${\cal B}$ on the self-dual boundary conditions,
we consider the $L_y=1$ strips with DBC2.  We shall denote this family
generically as $L$, and $L,1$ or $L1$ to specify the width.  The number of
$\lambda$'s in the form (\ref{zgsum}), is, from \cite{dg}, 
\beq
N_{Z,L,L_y,\lambda} = { 2L_y+1 \choose L_y+1} \ . 
\label{nztotdbc2}
\eeq
In particular, for $L_y=1$, this gives $N_{Z,L,1,\lambda}=3$.  We find 
\beq
\lambda_{L,1,1} = v \ , \quad 
\lambda_{L,1,2}=\lambda_{S,1,1} \ , \quad \lambda_{L,1,3}=\lambda_{S,1,2} \ . 
\label{lam123L1}
\eeq
The corresponding coefficients in (\ref{zgsum}) are 
\beq
c_{L1,1} = \kappa^{(2)}=q(q-2)
\label{c1L1}
\eeq
\beq
c_{L1,j} = \kappa^{(1)}=q \quad {\rm for} \quad j=2,3
\label{c23L1}
\eeq
where \cite{dg}
\beqs
\kappa^{(d)} & = & 2\Bigg [ U_{2d}\Big ( \frac{\sqrt{q}}{2} \Big ) -
T_{2d}\Big ( \frac{\sqrt{q}}{2} \Big ) \Bigg ] \cr\cr
& = & \sqrt{q} \ U_{2d-1} \Big ( \frac{\sqrt{q}}{2} \Big ) \cr\cr
& = & \sum_{j=0}^{d-1} (-1)^j { 2d-1-j \choose j} q^{d-j}
\label{kappad}
\eeqs
where $T_n(x)$ and $U_n(x)$ are the Chebyshev polynomials of the first and
second kinds.  Structural properties of $Z(G,q,v)$ for these strips have
interesting connections with Temperley-Lieb algebras and Bratteli diagrams, 
which were pointed out in \cite{dg}.  

\begin{figure}[hbtp]
\vspace{-10mm}
\centering
\leavevmode
\epsfxsize=3in
\begin{center}
\leavevmode
\epsffile{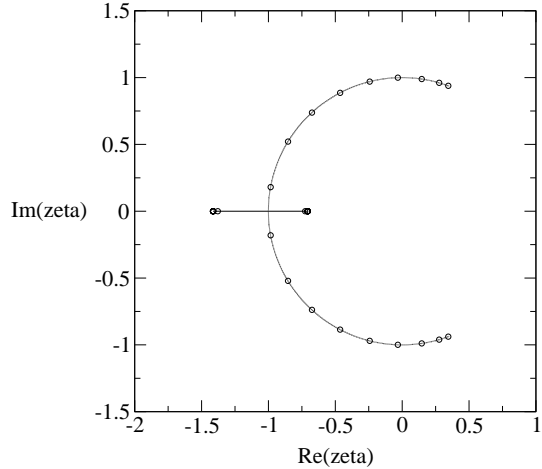}
\end{center}
\vspace{-10mm}
\caption{\footnotesize{Singular locus ${\cal B}$ in the $\zeta$ 
plane for the free energy of the $q=2$ Potts model on  
the $L_x \to \infty$ limit of the $L_y=1$ strip DBC2 boundary
conditions. For comparison, zeros of $Z$ for $L_x=20$ are shown.}}
\label{whpxy1vq2}
\end{figure}

\begin{figure}[hbtp]
\vspace{-10mm}
\centering
\leavevmode
\epsfxsize=3in
\begin{center}
\leavevmode
\epsffile{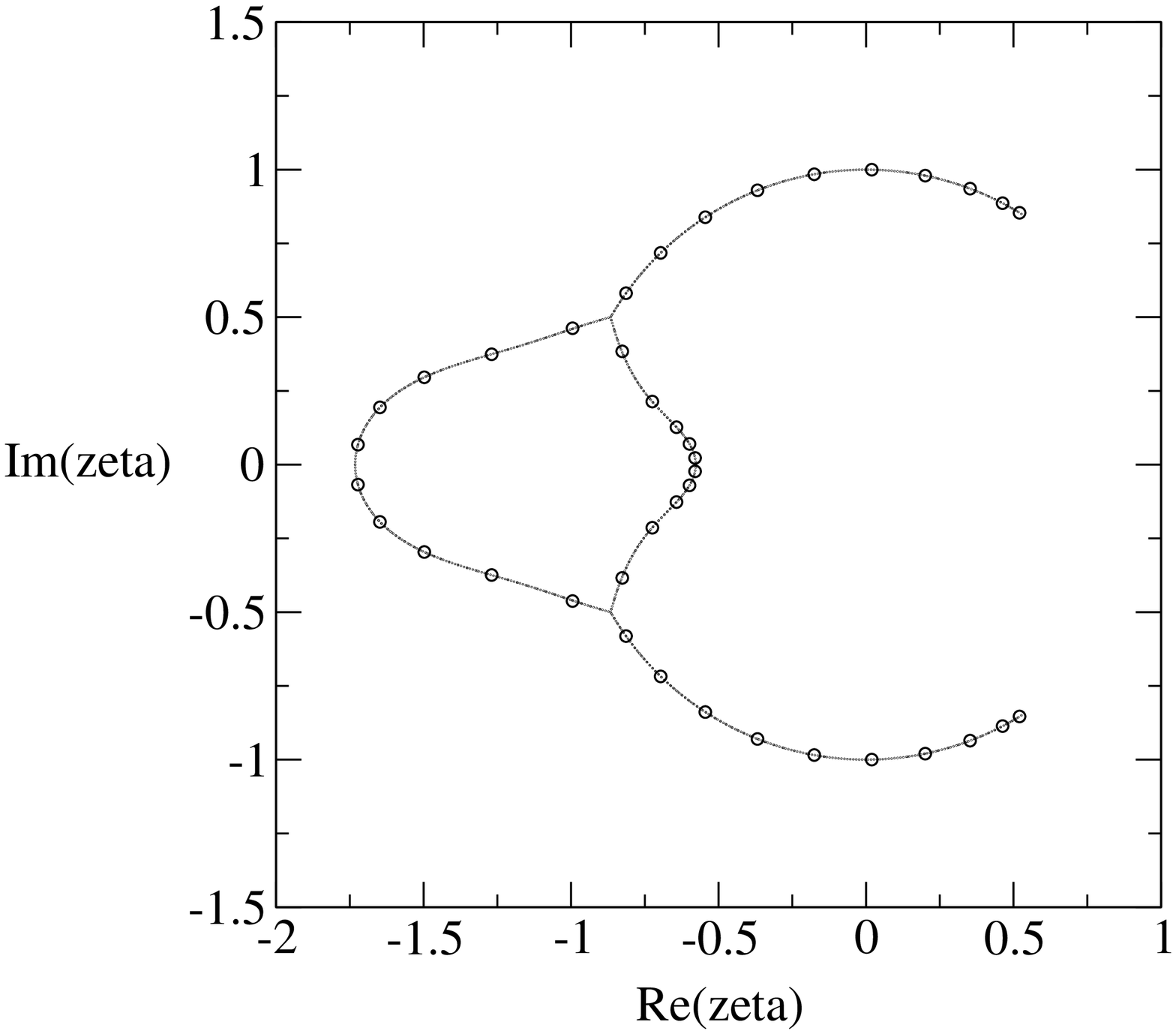}
\end{center}
\vspace{-10mm}
\caption{\footnotesize{Same as Fig. \ref{whpxy1vq2} for $q=3$.}}
\label{whpxy1vq3}
\end{figure}

\begin{figure}[hbtp]
\vspace{-10mm}
\centering
\leavevmode
\epsfxsize=3in
\begin{center}
\leavevmode
\epsffile{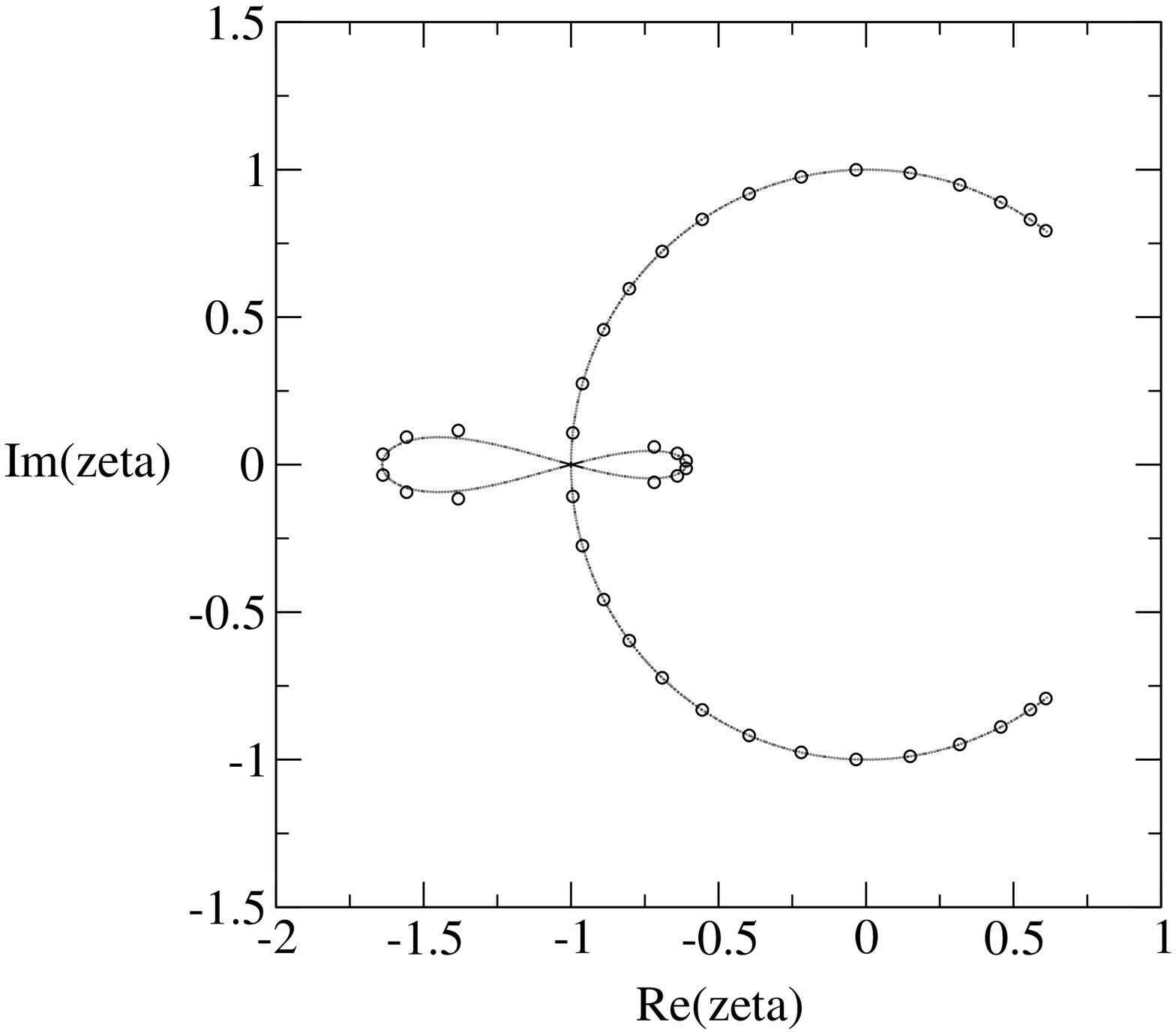}
\end{center}
\vspace{-10mm}
\caption{\footnotesize{Same as Fig. \ref{whpxy1vq2} for $q=4$.}}
\label{whpxy1vq4}
\end{figure}

\begin{figure}[hbtp]
\vspace{-10mm}
\centering
\leavevmode
\epsfxsize=3in
\begin{center}
\leavevmode
\epsffile{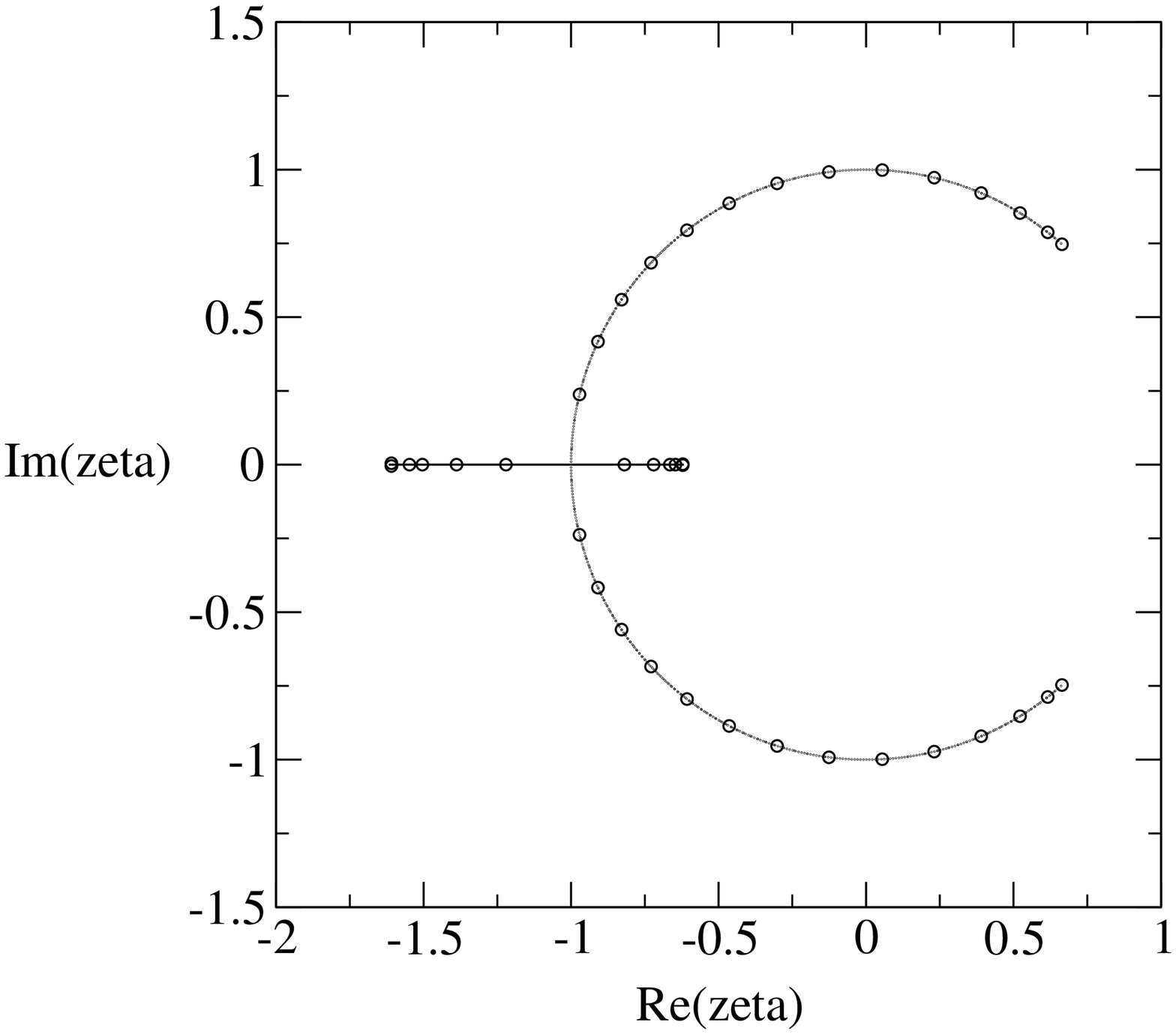}
\end{center}
\vspace{-10mm}
\caption{\footnotesize{Same as Fig. \ref{whpxy1vq2} for $q=5$.}}
\label{whpxy1vq5}
\end{figure}

\begin{figure}[hbtp]
\vspace{-10mm}
\centering
\leavevmode
\epsfxsize=3in
\begin{center}
\leavevmode
\epsffile{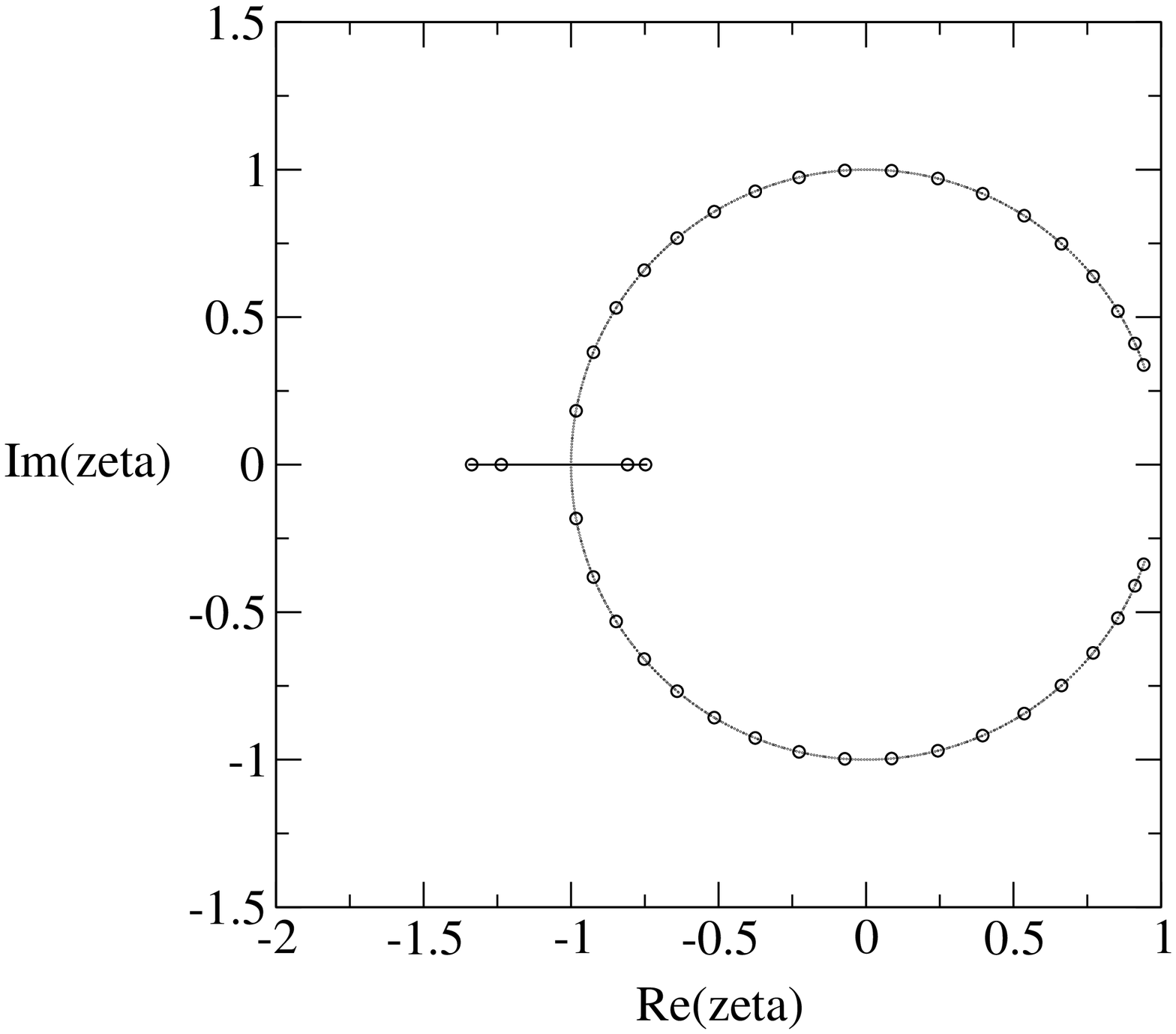}
\end{center}
\vspace{-10mm}
\caption{\footnotesize{Same as Fig. \ref{whpxy1vq2} for $q=100$.}}
\label{whpxy1vq100}
\end{figure}

In Figs. \ref{whpxy1vq2}-\ref{whpxy1vq100} we show the locus ${\cal B}$ and
associated complex-temperature phase diagram in the $\zeta$ plane for the
values $q=2$, 3, 4, 5, and 100.  (Note that the locus ${\cal B}$ shown in
Fig. \ref{whpxy1vq2} is ${\cal B}_{nq}$.)  The arc endpoints of the portion of
${\cal B}$ lying on the circle $|\zeta|=1$ are the same as for the ($L_x \to
\infty$ limit of the) $L_y=1$ strip with DBC1, i.e., $\zeta_{ae},\zeta_{ae}^*$
given in eq. (\ref{zetaae}).  The reason for this property is that in this area
of the $\zeta$ plane the locus ${\cal B}$ is determined by the equality in
magnitude of two terms, $\lambda_{L,1,2}$ and $\lambda_{L,1,3}$, which are
common to the partition functions for DBC1 and DBC2.  In our calculations for
$L_y=2,3$ we have found the same property to hold, so that for a given width
$L_y$ and value of $q$, for the strips considered here, the locations of the
right-hand arc endpoints are independent of whether one uses DBC1 or DBC2
boundary conditions.  However, there is an interval in $q$ for which
$\lambda_{L,1,1}$ is dominant in the vicinity of $\zeta=-1$, and this leads to
at least one complex-temperature O phase (in the nomenclature of \cite{chisq}).
Figs. \ref{whpxy1vq3} and \ref{whpxy1vq4} illustrate this for the cases $q=3$
and $q=4$.  One observes complex-conjugate triple points on ${\cal B}$ for
$q=3$.  An exactly solved case of a complex-temperature phase boundary
exhibiting such a triple point was given in \cite{z6}, where it was shown how
this triple point results from three curves on ${\cal B}$ coming together such
that as one travels along a given curve, beyond the intersection point the
$\lambda$'s that were leading and degenerate in magnitude on this curve are no
longer leading, so that their degeneracy is not relevant to ${\cal B}$.  In our
exact calculations of ${\cal B}$ in the $q$ plane for $n \to \infty$ limits of
chromatic polynomials we have found a number of such triple points (e.g.,
\cite{strip}-\cite{wcy},\cite{hs}).  Considering ${\cal B}$ as the union of the
various curves and line segments that comprise it, this is a multiple point
(intersection point) on ${\cal B}$ since it lies on multiple branches of ${\cal
B}$.  In a different nomenclature in which one considers each of the algebraic
curves individually, including the portions where the pairs of
degenerate-magnitude $\lambda$'s are not dominant so that these portions are
not on ${\cal B}$, then such triple points are not multiple points on each
individual algebraic curve, since these individual curves pass through the
triple point as shown in Fig. 2 in \cite{z6} (see also
\cite{ss00,biggscurves}).

 In general, as we did for the DBC1 strips, we find that for any finite $q$ no
matter how large, ${\cal B}$ deviates from the circle $|\zeta|=1$.  As
discussed above, the gap that opens in the circle in the vicinity of $\zeta=1$
is a property that is special to the quasi-one-dimensional nature of these
infinite-length, finite width strips.  However, just as our previous exact
results showed that certain complex-temperature properties of
quasi-one-dimensional spin models were similar to those of the same models in
2D \cite{is1d,a,ta,hca,ka,s3a}, so also the deviations in the region around
$\zeta=-1$ are indicative of what can happen in 2D in this case.  Note that for
$q > 1$, the point $\zeta=-1$, i.e., $v=-\sqrt{q}$, is a complex-temperature,
rather than physical, point.  We can now use our results to address the issue
of the radius of convergence of the $1/q$ expansion as regards the form of
${\cal B}$.  In general, if an expansion of some quantity in a variable
$\epsilon$ has a finite radius of convergence $\epsilon_c$, then, roughly
speaking, the behavior for $|\epsilon| < \epsilon_c$ should be qualitatively
the same as for $\epsilon=0$.  Our results suggest that, at least for the
infinite-length finite-width strips, the $1/q$ expansion has zero radius of
convergence insofar as properties of the locus ${\cal B}$ are concerned.  This
follows since ${\cal B}$ in the $\zeta$ plane differs qualitatively for any
nonzero value of $1/q$ from its form at $1/q=0$.  We know that the deviation
near $\zeta=1$, i.e. the physical PM-FM transition temperature given by
$v_c=\sqrt{q}$, that occurs for these quasi-one-dimensional strips will be
absent in 2D.  However, the deviation in the vicinity of the
complex-temperature point $\zeta=-1$ should be a more general feature, not
limited to the quasi-one-dimensional nature of the strips considered here.
This inference follows from (i) our previous experience comparing exactly
determined complex-temperature features of Ising model phase diagrams for
infinite-length strips and in 2D, (ii) the observed scatter of Fisher zeros in
2D \cite{mbook,chw,pfef} in this complex-temperature region, and (iii) reliable
determinations of locations of complex-temperature singularities via analyses
of low-temperature series \cite{pfef,hcl,p,p2} and the correlation of these
with points on ${\cal B}$ \cite{chisq,chitri,pfef,p,p2}.  Hence for the locus
${\cal B}$, our present exact results suggest that the $1/q$ expansion has zero
radius of convergence.  We emphasize that this does not reduce the value of
these large-$q$ expansions, since the point where the deviation occurs is
generically a complex-temperature point, and this type of deviation does not
occur near the physical PM-FM phase transition point.  Indeed, large-$q$
expansions yield excellent agreement with Monte Carlo measurements of
thermodynamic quantities in the Potts model \cite{lacaze,arisue99b} and also
for the ground state degeneracy $W(q)$ in the $T=0$ Potts antiferromagnet
\cite{w},\cite{nagle}-\cite{wn}.  Furthermore, it is easily checked that the
dominant term $\lambda$ that determines the free energy on the strips that we
consider does have a well-defined large-$q$ expansion (in the variable
$1/\sqrt{q}$). For example, for $L_y=1$ with DBC1 or DBC2, removing the leading
factor of $q$, we have the expansion for $q^{-1}\lambda_{S,1,1}$ for large $q$:
\beqs
q^{-1}\lambda_{S,1,1} & = & 1 + 2vq^{-1} + v^2(v+1)q^{-2} +
v^3(v^2-1)q^{-3} + O(q^{-4}) \cr\cr
                      & = & 1 + \frac{\zeta(\zeta^2-2)}{\zeta^2-1}q^{-1/2} 
-\frac{\zeta^2(\zeta^2+\zeta-1)(\zeta^2-\zeta-1)}{(\zeta^2-1)^3}q^{-1} \cr\cr
& & + \frac{\zeta^3(\zeta^2+1)(\zeta^2+\zeta-1)(\zeta^2-\zeta-1)}
{(\zeta^2-1)^5}q^{-3/2} + O(q^{-2})  \ .
\label{lams11largeq}
\eeqs
Note the poles at $\zeta=\pm 1$ in this expansion.  
Parenthetically, we note that our findings here concerning the large-$q$
expansion are not related to our earlier studies of families of graphs for
which $W(q)$ has no large-$q$ expansion \cite{w,wa,wa2,wa3,bcc} (see also
\cite{sokalzero}) since in those cases, the breakdown of the $1/q$ expansion is
equivalent to the property that the singular locus ${\cal B}$ is noncompact in
the $q$ plane, passing through the origin of the $1/q$ plane.  This sort of
breakdown does not occur for the present families of graphs, as is clear from
the fact that ${\cal B}$ is compact in the $q$ plane, shown in \cite{dg} and
above. 

\section{Wider Strips}

We have also calculated $Z(G,q,v)$ for arbitrary $q$ and $v$ for wider strips
with $L_y=2$ and $L_y=3$, for which our general formulas (\ref{nztotdbc1}) and
(\ref{nztotdbc2}) from \cite{dg} yield for the number of $\lambda$'s the
results $N_{Z,DBC1,2,\lambda}=5$, $N_{Z,DBC2,2,\lambda}=10$,
$N_{Z,DBC1,3,\lambda}=14$, and $N_{Z,DBC2,3,\lambda}=35$.  The analytic
expressions for the $\lambda$'s are too lengthy to list here.  In several
figures below we show plots of Fisher zeros in the $\zeta$ plane for various
values of $q$.

\begin{figure}[hbtp]
\vspace{-10mm}
\centering
\leavevmode
\epsfxsize=3in
\begin{center}
\leavevmode
\epsffile{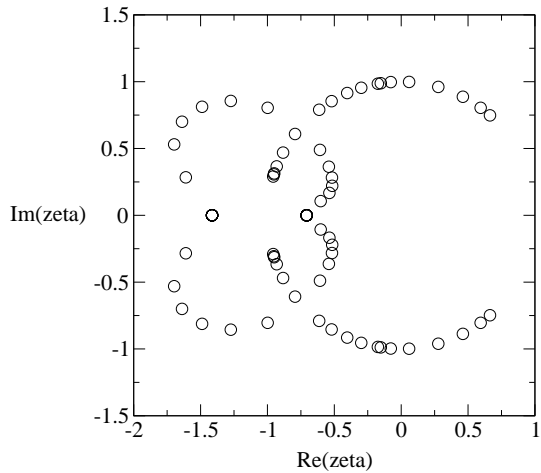}
\end{center}
\vspace{-10mm}
\caption{\footnotesize{Complex-temperature Fisher zeros in the $\zeta$
plane for the $q=2$ Potts model on the $L_y=2$ strip with DBC2 boundary
conditions with $L_x=20$ ($n=40$).}}
\label{whpxy2vq2}
\end{figure}

\begin{figure}[hbtp]
\vspace{-10mm}
\centering
\leavevmode
\epsfxsize=3in
\begin{center}
\leavevmode
\epsffile{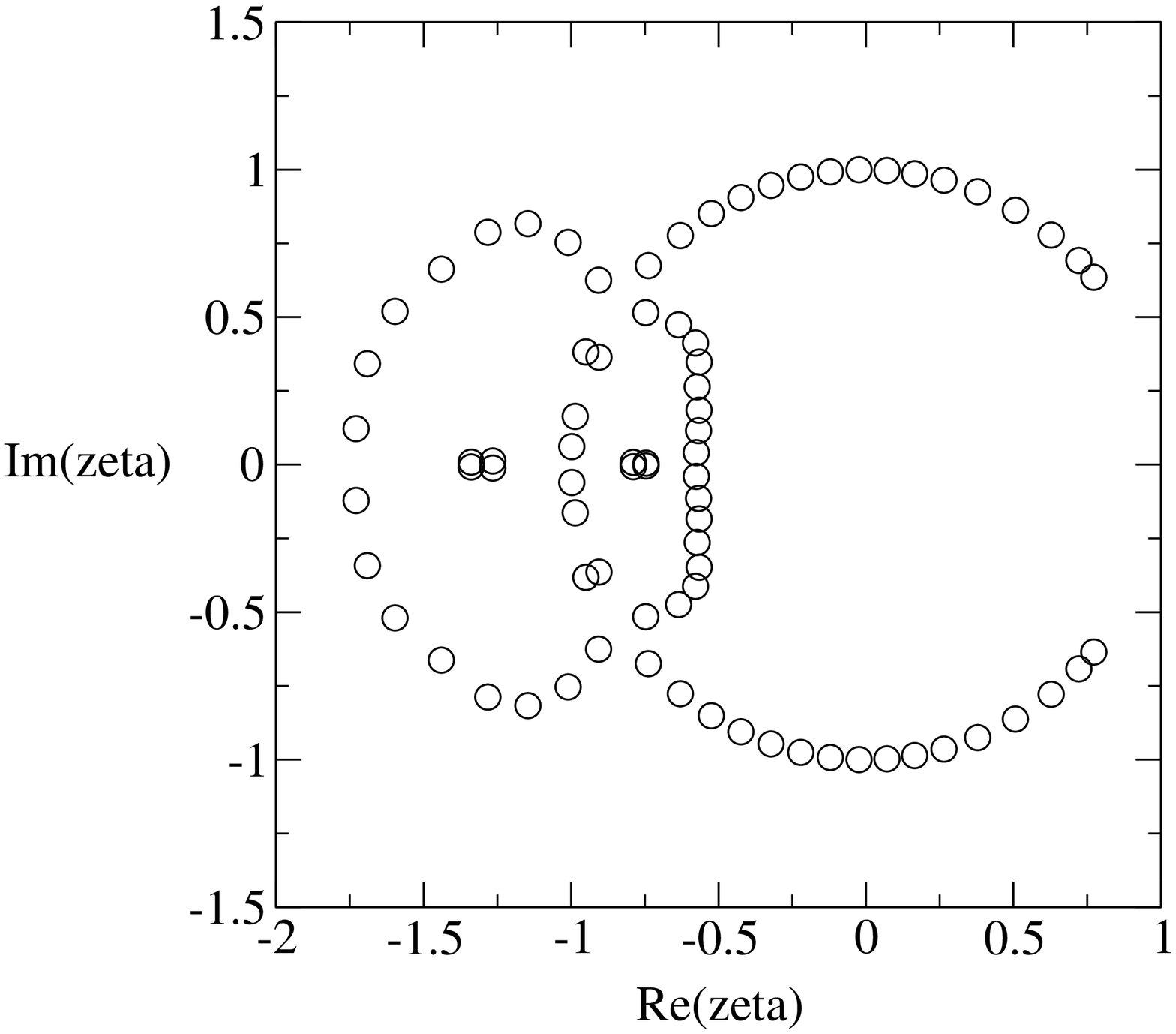}
\end{center}
\vspace{-10mm}
\caption{\footnotesize{Same as Fig. \ref{whpxy2vq2} for $q=3$.}}
\label{whpxy2vq3}
\end{figure}

\begin{figure}[hbtp]
\vspace{-10mm}
\centering
\leavevmode
\epsfxsize=3in
\begin{center}
\leavevmode
\epsffile{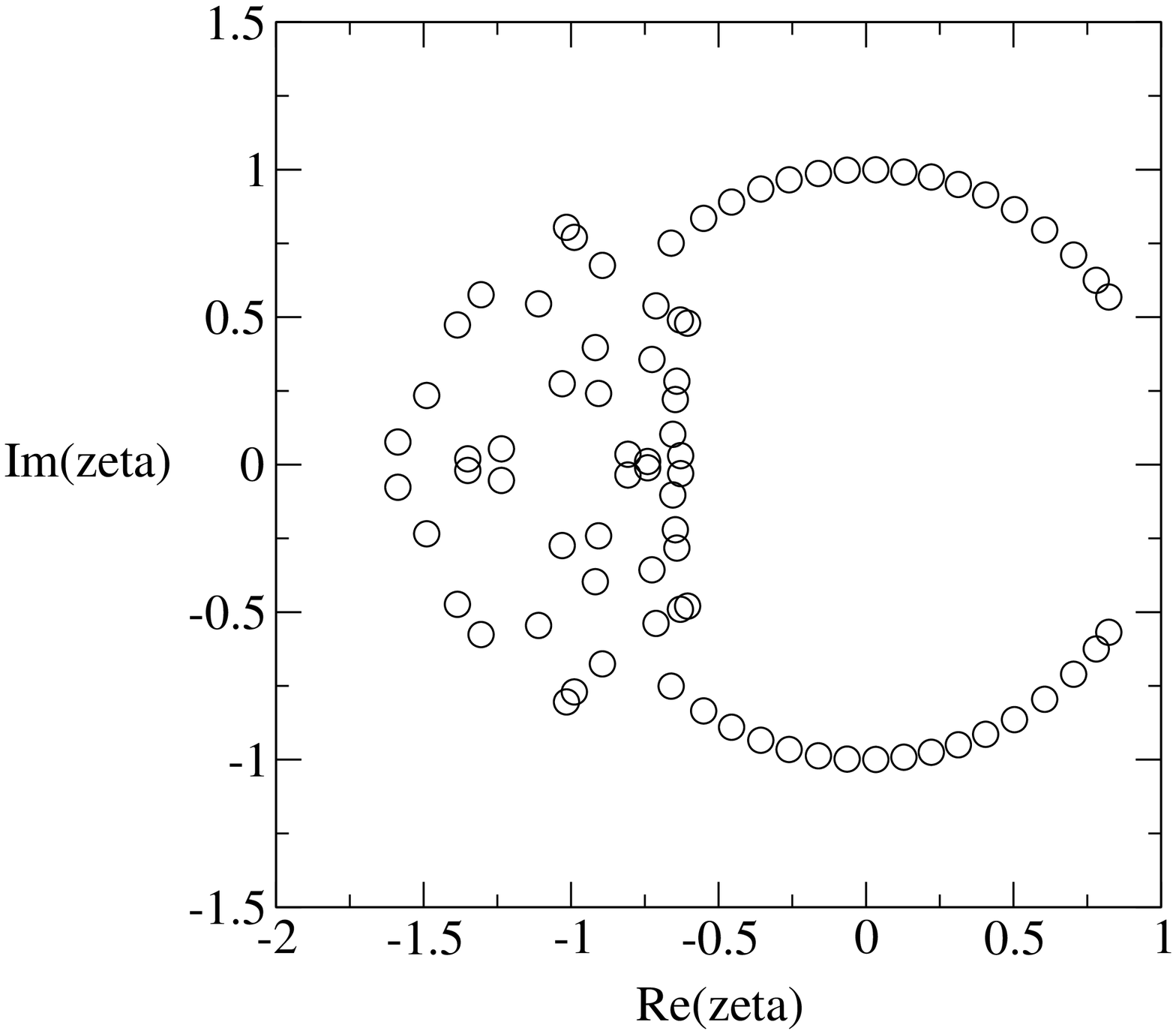}
\end{center}
\vspace{-10mm}
\caption{\footnotesize{Same as Fig. \ref{whpxy2vq2} for $q=4$.}}
\label{whpxy2vq4}
\end{figure}

\begin{figure}[hbtp]
\vspace{-10mm}
\centering
\leavevmode
\epsfxsize=3in
\begin{center}
\leavevmode
\epsffile{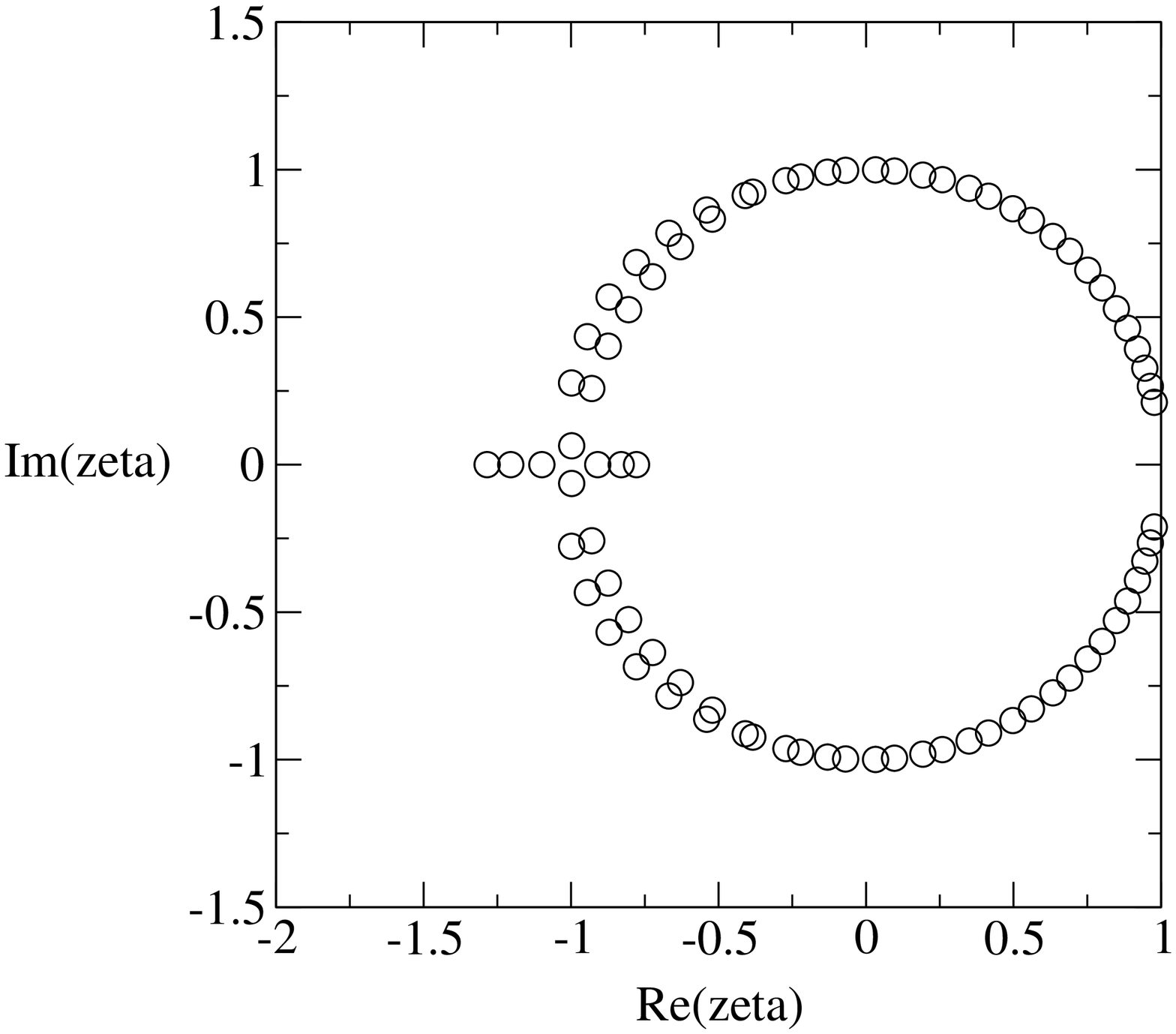}
\end{center}
\vspace{-10mm}
\caption{\footnotesize{Same as Fig. \ref{whpxy2vq2} for $q=100$.}}
\label{whpxy2vq100}
\end{figure}

\begin{figure}[hbtp]
\vspace{-10mm}
\centering
\leavevmode
\epsfxsize=3in
\begin{center}
\leavevmode
\epsffile{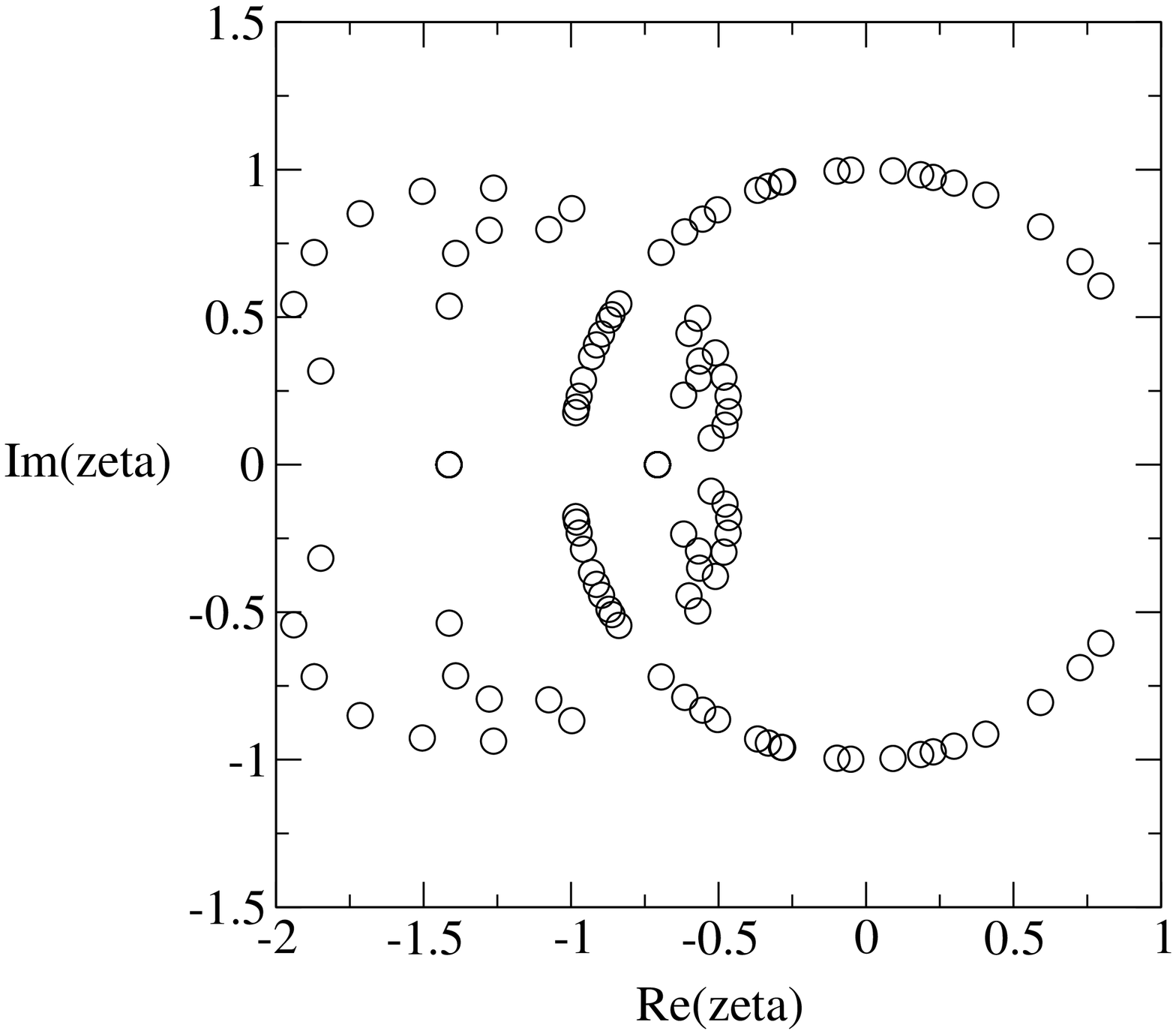}
\end{center}
\vspace{-10mm}
\caption{\footnotesize{Complex-temperature Fisher zeros in the $\zeta$
plane for the $q=2$ Potts model on the $L_y=3$ strip with DBC2 boundary
conditions with $L_x=20$ ($n=60$).}}
\label{whpxy3vq2}
\end{figure}

\begin{figure}[hbtp]
\vspace{-10mm}
\centering
\leavevmode
\epsfxsize=3in
\begin{center}
\leavevmode
\epsffile{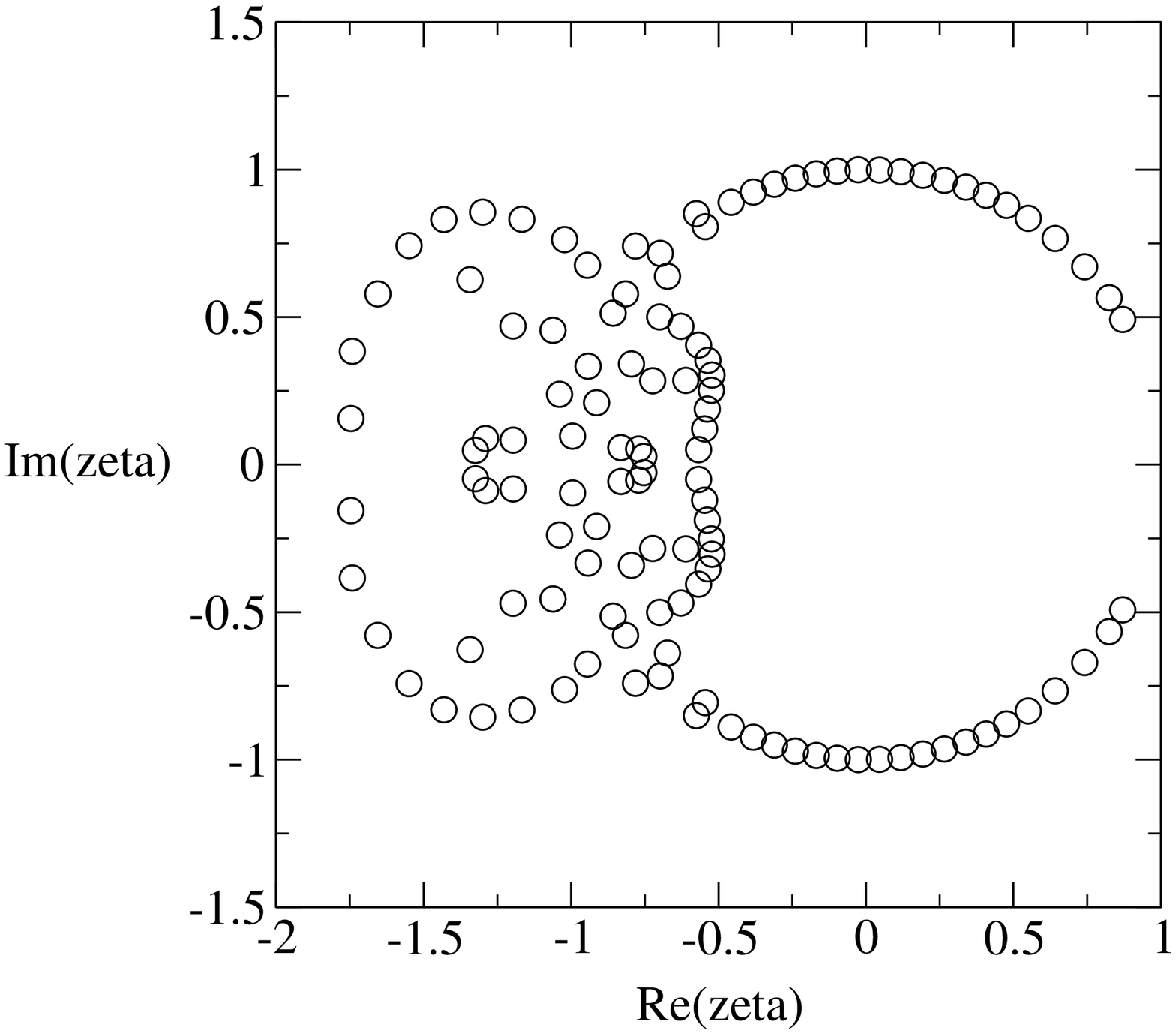}
\end{center}
\vspace{-10mm}
\caption{\footnotesize{Same as Fig. \ref{whpxy3vq3} for $q=3$.}}
\label{whpxy3vq3}
\end{figure}

\begin{figure}[hbtp]
\vspace{-10mm}
\centering
\leavevmode
\epsfxsize=3in
\begin{center}
\leavevmode
\epsffile{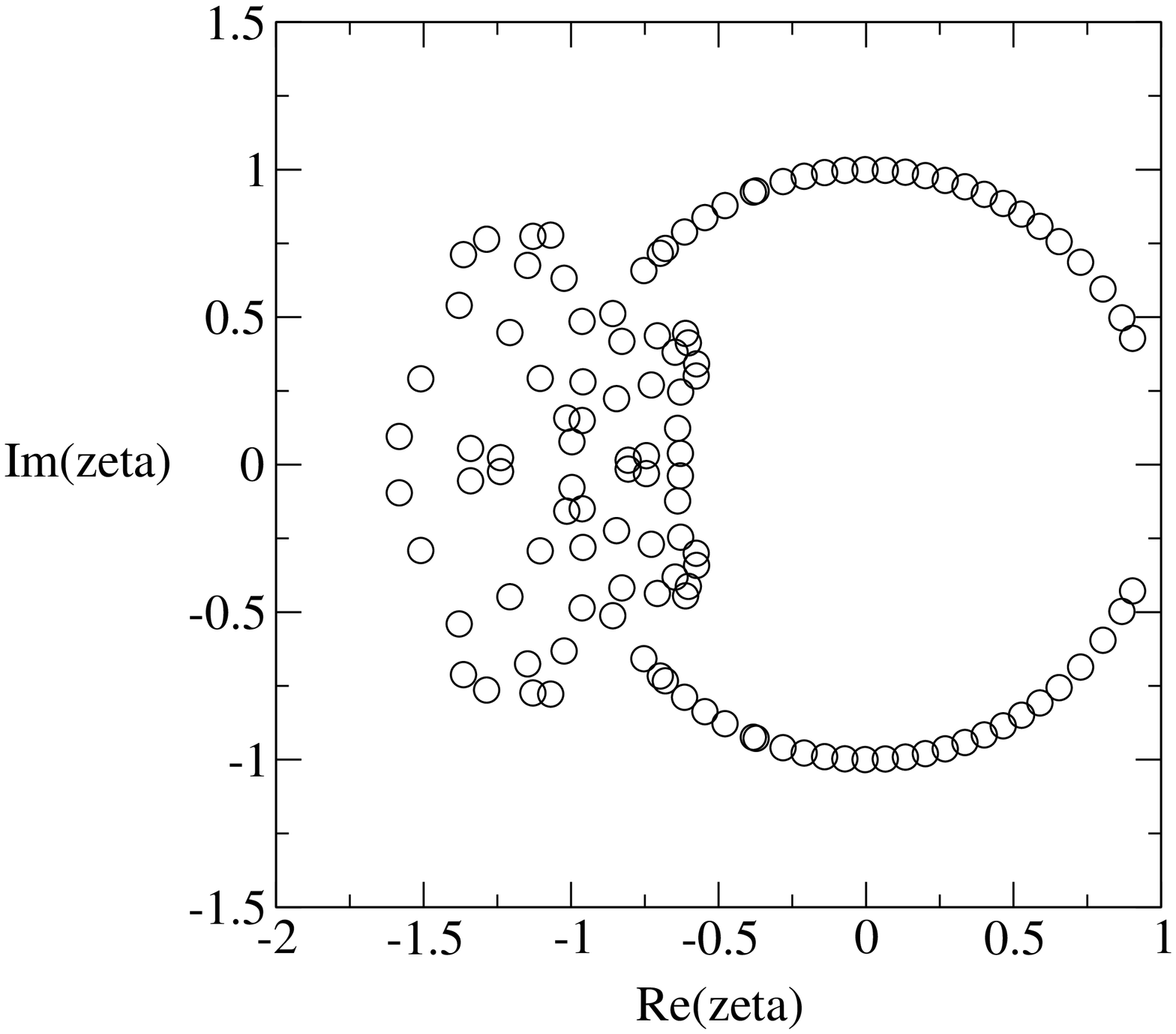}
\end{center}
\vspace{-10mm}
\caption{\footnotesize{Same as Fig. \ref{whpxy3vq4} for $q=4$.}}
\label{whpxy3vq4}
\end{figure}

\begin{figure}[hbtp]
\vspace{-10mm}
\centering
\leavevmode
\epsfxsize=3in
\begin{center}
\leavevmode
\epsffile{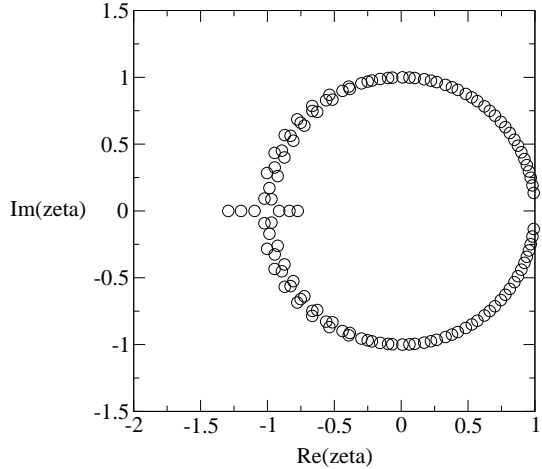}
\end{center}
\vspace{-10mm}
\caption{\footnotesize{Same as Fig. \ref{whpxy3vq2} for $q=100$.}}
\label{whpxy3vq100}
\end{figure}

 It is interesting to note that the simplification of the complex-temperature
phase diagram for the Potts model to the circle $|\zeta|=1$ for $q \to \infty$
proved in \cite{wuetal} and studied further here with exact results is somewhat
similar to the simplification of the complex-temperature phase diagram for the
2D spin-$s$ Ising model in the limit $s \to \infty$ to the circle $|u_s|=1$ in
the plane of the Boltzmann variable $u_s=e^{-K/s^2}$ \cite{guts,hsising,is1d}.
In both cases, the approach to the limit is singular in the sense that there
are deviations from the asymptotic locus.

\section{Conclusions}

In summary, we have presented exact calculations of the Potts model partition
function $Z(G,q,v)$ on self-dual strip graphs $G$ of the square lattice with
fixed width $L_y$ and arbitrarily great length $L_x$ with two types of boundary
conditions.  In the infinite-length limit we have studied the resultant
complex-temperature phase diagram. In particular, we have considered the widths
$L_y=1,2,3$.  We have used these results to study the approach to the large-$q$
limit of ${\cal B}$, where this locus is the unit circle $|\zeta|=1$.  We find
that, for a given $L_y$ and set of self-dual boundary conditions, a portion of
${\cal B}$ lies on this unit circle, while for any finite $q$, a portion
deviates from this circle.  For the strips considered here we find that the
right-hand arc endpoints on the circle are independent of whether the
longitudinal boundary conditions are periodic or free and to curve around and
pinch the real axis at $\zeta=1$ as $q \to \infty$.  For a fixed value of $q$,
as $L_y$ increases, the right-hand arc endpoints move closer to $\zeta=1$.  On
the left, the nature of the complex-temperature phase diagram was found to
depend in detail on both the type of boundary conditions and the width of the
strip.  As $q \to \infty$, the deviations typically include real line segments
as well as possible O phases.  One feature was found for each width considered,
namely that for the DBC2 strips, for $q=3$, ${\cal B}$ crosses the $\zeta$ at
$\zeta=-1/\sqrt{3}$ and at $-\sqrt{3}$.  We showed that this is equivalent to
the fact that for $v=-1$, each of the infinite-length strips that we studied in
\cite{dg} with $1 \le L_y \le 4$ and DBC2 had $q_c=3$ as for the infinite
square lattice.  As discussed, the gap in the locus ${\cal B}$ that opens
around $\zeta=1$ as $1/q$ increases above zero is a consequence of the
quasi-one-dimensional nature of the strips.  However, the behavior in the
$Re(\zeta) < 0$ region near $\zeta=-1$ can give some insight as to how ${\cal
B}$ could behave for large $q$ on the infinite square lattice.

Acknowledgment: This research was partially supported by the NSF grant 
PHY-97-22101.

\section{Appendix}

\subsection{General} 

The most compact way to express a Potts model partition function $Z(G,q,v)$ is
often in terms of the corresponding Tutte polynomial $T(G,x,y)$.  For the
reader's convenience, we recall the definition of the Tutte polynomial and some
basic formulas relating these functions here (e.g., \cite{a}).
For an arbitrary graph $G$ the Tutte polynomial of $G$, $T(G,x,y)$,
is given by \cite{tutte1}-\cite{tutte3}
\beq 
T(G,x,y)=\sum_{G^\prime \subseteq 
G} (x-1)^{k(G^\prime)-k(G)} (y-1)^{c(G^\prime)}
\label{tuttepol}
\eeq
where the spanning subgraph $G^\prime$ was defined in the introduction, and we
recall that $k(G^\prime)$, $e(G^\prime)$, and $n(G^\prime)=n(G)$ denote the
number of components, edges, and vertices of $G^\prime$, where 
\beq
c(G^\prime) = e(G^\prime)+k(G^\prime)-n(G^\prime)
\label{ceq}
\eeq
is the number of independent circuits in $G^\prime$. 
As stated in the text, $k(G)=1$ for the graphs of interest here.  Now let
\beq
x=1+\frac{q}{v} \ , \quad y=a=v+1
\label{xydef}
\eeq
so that 
\beq
q=(x-1)(y-1) \ .  
\label{qxy}
\eeq
Then
\beq
Z(G,q,v)=(x-1)^{k(G)}(y-1)^{n(G)}T(G,x,y) \ .
\label{ztutte}
\eeq

For a planar graph $G$ the Tutte polynomial satisfies the duality relation
\beq
T(G,x,y) = T(G^*,y,x)
\label{tuttesym}
\eeq
where $G^*$ is the (planar) dual to $G$.  As discussed in \cite{a}, 
the Tutte polynomial for recursively defined graphs comprised of $m$ 
repetitions of some subgraph has the form
\beq
T(G_m,x,y) = \sum_{j=1}^{N_\lambda} c_{T,G,j}(\lambda_{T,G,j})^m \ .
\label{tgsum}
\eeq
One special case of the Tutte polynomial of particular interest is the
chromatic polynomial $P(G,q)$. This is obtained by setting $y=0$,
i.e., $v=-1$, so that $x=1-q$; the correspondence is $P(G,q) =
(-q)^{k(G)}(-1)^{n}T(G,1-q,0)$.

\subsection{Strips with DBC1} 

The generating function representation for the Tutte polynomial for the strip
$S_m$ of the square lattice with length $L_x=m+1$ vertices, i.e., $m$ edges 
in each horizontal row, and of width $L_y$, with duality preserving boundary
conditions of type 1, is 
\beq
\Gamma_T(S_m,L_y,x,y,z) = \sum_{m=0}^\infty T(S_m,L_y,x,y)z^m \ . 
\label{gammatfbc}
\eeq
We have
\beq
\Gamma_T(S,L_y,x,y;z) = \frac{{\cal N}_T(S,L_y,x,y,z)}{{\cal
D}_T(S,L_y,x,y,z)} \ .
\label{gammas}
\eeq
For $L_y=1$ we find 
\beq
{\cal N}_T(S,1,x,y,z)=(x+y)-xyz
\label{numts}
\eeq
\beqs
{\cal D}_T(S,1,x,y,z) & = & 1 - (1+x+y)z + xyz^2 \cr\cr
                    & = & \prod_{j=1}^2 (1-\lambda_{T,S,1,j}z)
\label{dents}
\eeqs
with
\beq
\lambda_{T,S,1,(1,2)} = \frac{1}{2}\biggl [ 1+x+y \pm 
\Bigl (1+2(x+y)+(x-y)^2 \Bigr )^{1/2} \biggr ] \ . 
\label{lamstut12}
\eeq
\beq
T(S_m,x,y)=\biggl [
\frac{A_{T,S,0}\lambda_{T,S,1}+A_{T,S,1}}{\lambda_{T,S,1}-\lambda_{T,S,2}}
\biggr ] (\lambda_{T,S,1})^m + \biggl [
\frac{A_{T,S,0}\lambda_{T,S,2}+A_{T,S,1}}
{\lambda_{T,S,2}-\lambda_{T,S,1}} \biggr ] (\lambda_{T,S,2})^m \ .
\label{tssumform}
\eeq

For $L_y=2$ we find
\beq
{\cal N}_T(S,2,x,y,z)=A_{S2,0}+A_{S2,1}z+A_{S2,2}z^2+A_{S2,3}z^3+A_{S2,4}z^4
\label{numts2}
\eeq
where
\beq
A_{S2,0} = x+y+xy+x^2+y^2 
\eeq
\beq
A_{S2,1} = -\bigg [x+y+2(x^2+y^2)+3xy+5xy(x+y)+xy(x^2+y^2)+x^3+y^3+(xy)^2 
\bigg ]  
\eeq
\beq
A_{S2,2}=xy\bigg [3(x+y)+4(x^2+y^2)+6xy+3xy(x+y)+x^3+y^3+(xy)^2 \bigg ] 
\eeq
\beq
A_{S2,3} = -(xy)^2\bigg [2(x+y)+x^2+y^2+3xy+xy(x+y) \bigg ] 
\eeq
\beq
A_{S2,4} = (xy)^4 
\eeq
\beq
{\cal D}_T(S,2,x,y,z) = 1+b_{S2,1}z+b_{S2,2}z^2+b_{S2,3}z^3+b_{S2,4}z^4+
b_{S2,5}z^5
\label{dents2}
\eeq   
where
\beq
b_{S2,1} = -\bigg [3(1+x+y)+xy+x^2+y^2 \bigg ]
\eeq
\beq
b_{S2,2} = 1+3(x+y)+3(x^2+y^2)+8xy+(x^3+y^3)+5xy(x+y)+xy(x^2+y^2)+(xy)^2
\eeq
\beq
b_{S2,3} = -xy\bigg [3+5(x+y)+4(x^2+y^2)+6xy+(x^3+y^3)+3xy(x+y)+(xy)^2 \bigg ]
\eeq
\beq
b_{S2,4} = (xy)^2(1+x)(1+y)(1+x+y)
\eeq
\beq
b_{S2,5} = -(xy)^4 \ .
\eeq

For $L_y=3$ our general results in \cite{dg} yield the result that there are 14
terms, and we find that the $\lambda_{T,S,3,j}$'s are roots of an algebraic
equation of degree 14.  This is too lengthy to record here, but is available
upon request. 

\subsection{Strips with DBC2} 

We have 
\beq
T(L_m,L_y,x,y) = \sum_{j=1}^{N_{T,L,L_y,\lambda}} c_{T,L,L_y,j}
(\lambda_{T,L,L_y,j})^m
\label{tlxy}
\eeq
where $N_{T,L,L_y,\lambda}=N_{Z,L,L_y,\lambda}$, and our general formula for
this number, from \cite{dg} was given in (\ref{nztotdbc2}). 

\subsubsection{$L_y=1$}

We have $N_{T,L,1,\lambda}=3$ and 
\beq
\lambda_{T,L,1,1} = 1
\label{lamtutdbc21}
\eeq
\beq
\lambda_{T,L,1,j}=\lambda_{T,S,1,j-1} \quad {\rm for} \ j=2,3
\label{lamtutL123}
\eeq
where $\lambda_{T,S,1,1}$ and $\lambda_{T,S,1,2}$ were given in
eq. (\ref{lamstut12}). 
The corresponding coefficients are 
\beq
 c_{T,L,1,1} = q^{-1}\kappa^{(2)}=q-2=xy-x-y-1
\label{c1tutt}
\eeq
\beq
 c_{T,L,j} = q^{-1}\kappa^{(1)}=1 \quad {\rm for} \quad j=2,3
\label{c234tutt}
\eeq
where $\kappa^{(d)}$ was defined in eq. (\ref{kappad}) in the text.

\subsubsection{$L_y=2$}

We have $N_{T,L,2,\lambda}=10$.  The term $\lambda_{T,L,2,1}$ is 
\beq
\lambda_{T,L,2,1} = 1 \ .
\label{lamtutL2_1}
\eeq
The $\lambda_{T,L,2,j}$ for $2 \le j \le 5$ are solutions to the equation 
\beqs
& & \xi^4 - \Big [2(x+y)+3 \Big ]\xi^3 +\Big [x^2+y^2+3(x+y)+4xy+1 \Big ]\xi^2
 \cr\cr
& & - xy\Big [2(x+y)+3 \Big ]\xi + (xy)^2 = 0 \ .
\label{lamtutL2eq1}
\eeqs
The $\lambda_{T,L,2,j}$ for $6 \le j \le 10$ are solutions to the equation
\beq
\xi^5+b_{S2,1}\xi^4+b_{S2,2}\xi^3+b_{S2,3}\xi^2+b_{S2,4}\xi+b_{S2,5}=0 \ .
\label{lamtutL2eq2}
\eeq
The corresponding coefficients are 
\beq
c_{T,L,2,1} = q^{-1}\kappa^{(3)} = (q-1)(q-3) = (xy-x-y)(xy-x-y-2)
\label{c1tuttL2}
\eeq
\beq
c_{T,L,2,j} = q^{-1}\kappa^{(2)} = q-2 = xy-x-y-1 \quad {\rm for} 
\quad 2 \le j \le 5
\label{c25tuttL2}
\eeq
\beq
c_{T,L,2,j} = q^{-1}\kappa^{(1)} = 1 \quad {\rm for} \quad 6 \le j \le 10
\ .
\label{c610tuttL2}
\eeq

\subsubsection{$L_y=3$}

Here our general formulas in \cite{dg} yield the results that there are 
35 terms in all, comprised of (i) one term with coefficient 
$q^{-1}\kappa^{(4)}$, namely, $\lambda_{T,L,3,1}=1$, (ii) six terms
$\lambda_{T,L,3,j}$, $2 \le j \le 7$, with coefficient 
$q^{-1}\kappa^{(3)}$, (iii) 14 terms $\lambda_{T,L,3,j}$ for $8 \le j \le 21$
with coefficient $q^{-1}\kappa^{(2)}$, and (iv) 14 terms $\lambda_{T,L,3,j}$
with $22 \le j \le 35$ with coefficient $q^{-1}\kappa^{(1)}$.  The terms in
(ii) are roots of the sixth-degree equation
\beqs
& & \xi^6-(3x+3y+5)\xi^5+(3x^2+9yx+10x+3y^2+10y+6)\xi^4 \cr\cr
& & -(x^3+9yx^2+5x^2+9y^2x+20yx+6x+y^3+5y^2+6y+1)\xi^3 \cr\cr
& & +xy(3x^2+9yx+10x+3y^2+10y+6)\xi^2-x^2y^2(3x+3y+5)\xi+x^3y^3 = 0 \ .
\label{eqsix}
\eeqs
The equation of degree 14 for the $\lambda_{T,L,3,j}$ with coefficient 
$q^{-1}\kappa^{(1)}$ is the same as the single degree-14 equation for the terms
in the $L_y=3$ strip with DBC1.  Both this and the other degree-14 equation are
too lengthy to list here, but can be provided at request.

\subsection{Special Values of Tutte Polynomials}

For a given graph $G=(V,E)$, at certain special values of the arguments $x$ and
$y$, the Tutte polynomial $T(G,x,y)$ yields quantities of basic graph-theoretic
interest \cite{tutte3}-\cite{boll}.  We recall some definitions: a spanning
subgraph was defined at the beginning of the paper; a tree is a 
connected graph with no cycles; a forest is a graph containing one or 
more trees; and a spanning tree is a spanning subgraph that is a tree.  We 
recall that the graphs $G$ that we consider are connected.  Then the number 
of spanning trees of $G$, $N_{ST}(G)$, is
\beq
N_{ST}(G)=T(G,1,1) \ ,
\label{t11}
\eeq
the number of spanning forests of $G$, $N_{SF}(G)$, is
\beq
N_{SF}(G)=T(G,2,1) \ ,
\label{t21}
\eeq
the number of connected spanning subgraphs of $G$, $N_{CSSG}(G)$, is
\beq
N_{CSSG}(G)=T(G,1,2) \ ,
\label{T12}
\eeq
and the number of spanning subgraphs of $G$, $N_{SSG}(G)$, is
\beq
N_{SSG}(G)=T(G,2,2) \ .
\label{t22}
\eeq
Since the graphs that we consider are self-dual, and using the symmetry
relation (\ref{tuttesym}), we have 
\beq
N_{SF}(G)=N_{CSSG}(G^*)=N_{CSSG}(G) \ . 
\label{t21t12}
\eeq

 From our calculations of Tutte polynomials, we find the following results. 

\subsection{$L_y=1$, DBC1}

\beq
N_{ST}(S1_m)=\Bigg (1+\frac{2\sqrt{5}}{5} \ \Bigg )
\Bigg (\frac{3+\sqrt{5}}{2} \ \Bigg )^m +
\Bigg (1-\frac{2\sqrt{5}}{5} \ \Bigg )
\Bigg (\frac{3-\sqrt{5}}{2} \ \Bigg )^m 
\label{nst_s1m}
\eeq
\beq
N_{SF}(S1_m)=N_{CSSG}(S1_m)=\Bigg ( \frac{3}{2}+\sqrt{2} \Bigg )
 (2+\sqrt{2} \ )^m + \Bigg ( \frac{3}{2}-\sqrt{2} \Bigg )(2-\sqrt{2} \ )^m
\label{nsf_s1m}
\eeq
\beq
N_{SSG}(S1_m)=2^{e(S1_m)}=2^{2(m+1)} \ .
\label{nssg_s1m}
\eeq

Since these quantities grow exponentially, it is natural to define an
associated quantity that measures this growth \cite{wu77,wust,sw}.  
In particular, for the number of spanning trees, we define 
\beq
z_{\{G\}} = \lim_{n \to \infty} n^{-1} \ln N_{ST(\{G\})} \ .
\label{zg}
\eeq
For the present $L_y=1$, DBC1 strips, we thus have
$z=\ln[(3+\sqrt{5} \ )/2] \simeq 0.9624$. A general upper bound on the number
of spanning trees of a graph $G$ is \cite{grim}
\beq
N_{ST}(G) \le \frac{1}{n}\biggl ( \frac{2|E|}{n-1}\biggr )^{n-1},
\label{kboundg}
\eeq
For the present $L_y=1$, DBC1 strips, this gives the upper bound 
$z < 2\ln 2 \simeq 1.386$, which is seen to be satisfied by our result. 

\subsection{$L_y=1$, DBC2}

\beq
N_{ST}(L1_m)=-2+\Bigg (\frac{3+\sqrt{5}}{2} \ \Bigg )^m +
\Bigg ( \frac{3+\sqrt{5}}{2} \ \Bigg )^m
\label{nst_l1m}
\eeq
\beq
N_{SF}(L1_m)=N_{CSSG}(L1_m)=-2+(2+\sqrt{2} \ )^m + (2-\sqrt{2} \ )^m
\label{nsf_l1m}
\eeq
\beq
N_{SSG}(L1_m)=2^{e(L1_m)}=2^{2m} \ .
\label{nssg_l1m}
\eeq
As discussed before \cite{a,sw}, for a given $L_y$ and set of transverse
boundary conditions, the value of $z$ is the same, independent of whether the
longitudinal boundary conditions are free, as in DBC1 or periodic, as for
DBC2. 

\subsection{$L_y=2$, DBC2}

\beq
N_{ST}(L2_m)= 3-2 \sum _{j=2}^5 (\lambda_{T,L,2,j}(1,1)\ )^m +
\sum _{j=6}^{10} (\lambda_{T,L,2,j}(1,1) \ )^m
\label{nst_l2m}
\eeq
where $\lambda_{T,L,2,j}(1,1)$, $2 \le j \le 5$ and $6 \le j \le 10$ are
the roots of eq. (\ref{lamtutL2eq1}) and eq. (\ref{lamtutL2eq2}),
respectively, for $x=1,y=1$, viz., $\xi^4-7\xi^3+13\xi^2-7\xi+1=0$,
and $(\xi-1)(\xi^4-11\xi^3+25\xi^2-11\xi+1)=0$.
\beqs
N_{SF}(L2_m) = N_{CSSG}(L2_m) & = & 3 - 2 \Biggl [ \Bigl (
\frac{3+\sqrt{5}}{2} \ \Bigr )^m + \Bigl ( \frac{3-\sqrt{5}}{2} \ \Bigr
)^m + (3+\sqrt{5} \ )^m \cr\cr
& + & (3-\sqrt{5} \ )^m \Biggr ] + \sum_{j=6}^{10}
(\lambda_{T,L,2,j}(2,1)\ )^m
\label{nsf_l2m}
\eeqs
where $\lambda_{T,L,2,j}(2,1)$, $6 \le j \le 10$ are the roots of
eq. (\ref{lamtutL2eq2}) for $x=2,y=1$, viz.,
$\xi^5-19\xi^4+94\xi^3-162\xi^2+96\xi-16=0$.
\beq
N_{SSG}(L2_m)=2^{e(L2_m)}=2^{4m} \ .
\label{nssg_l2m}
\eeq
Hence, in particular, for spanning trees, we find $z \simeq 1.044$ for
$L_y=2$. It is straightforward to use our exact calculations of Tutte
polynomials for $L_y=3$ with DBC1 and DBC2 boundary conditions to list similar
results for spanning trees, etc.

\vfill
\eject
\end{document}